\documentclass[aps,prc,twocolumn,groupedaddress,showpacs	]{revtex4-1}
\usepackage{graphicx}
\usepackage{verbatim}
\usepackage{color}
\usepackage{amsmath}

\begin{document}


\title{Associate $\mathrm{K^0}$ production in p+p collisions at 3.5\,GeV: The role of $\mathrm{\Delta(1232)^{++}}$}


\author{G.~Agakishiev$^{7}$, O.~Arnold$^{9,10}$, D.~Belver$^{18}$, A.~Belyaev$^{7}$, 
J.C.~Berger-Chen$^{9,10,*}$, A.~Blanco$^{2}$, M.~B\"{o}hmer$^{10}$, J.~L.~Boyard$^{16}$, P.~Cabanelas$^{18}$, 
S.~Chernenko$^{7}$, A.~Dybczak$^{3}$, E.~Epple$^{9,10}$, L.~Fabbietti$^{9,10}$, O.~Fateev$^{7}$, 
P.~Finocchiaro$^{1}$, P.~Fonte$^{2,b}$, J.~Friese$^{10}$, I.~Fr\"{o}hlich$^{8}$, T.~Galatyuk$^{5,c}$, 
J.~A.~Garz\'{o}n$^{18}$, R.~Gernh\"{a}user$^{10}$, K.~G\"{o}bel$^{8}$, M.~Golubeva$^{13}$, D.~Gonz\'{a}lez-D\'{\i}az$^{5}$, 
F.~Guber$^{13}$, M.~Gumberidze$^{5}$, T.~Heinz$^{4}$, T.~Hennino$^{16}$, R.~Holzmann$^{4}$, 
A.~Ierusalimov$^{7}$, I.~Iori$^{12,e}$, A.~Ivashkin$^{13}$, M.~Jurkovic$^{10}$, B.~K\"{a}mpfer$^{6,d}$, 
T.~Karavicheva$^{13}$, I.~Koenig$^{4}$, W.~Koenig$^{4}$, B.~W.~Kolb$^{4}$, G.~Kornakov$^{5}$, 
R.~Kotte$^{6}$, A.~Kr\'{a}sa$^{17}$, F.~Krizek$^{17}$, R.~Kr\"{u}cken$^{10}$, H.~Kuc$^{3,16}$, 
W.~K\"{u}hn$^{11}$, A.~Kugler$^{17}$, T.~Kunz$^{10}$, A.~Kurepin$^{13}$, V.~Ladygin$^{7}$, R.~Lalik$^{9}$, 
K.~Lapidus$^{9,10,*}$, A.~Lebedev$^{14}$, L.~Lopes$^{2}$, 
M.~Lorenz$^{8}$, L.~Maier$^{10}$, A.~Mangiarotti$^{2}$, J.~Markert$^{8}$, V.~Metag$^{11}$, 
J.~Michel$^{8}$, C.~M\"{u}ntz$^{8}$, R.~M\"{u}nzer$^{9,10}$,L.~Naumann$^{6}$, Y.~C.~Pachmayer$^{8}$, 
M.~Palka$^{3}$, Y.~Parpottas$^{15,f}$, V.~Pechenov$^{4}$, O.~Pechenova$^{8}$, J.~Pietraszko$^{4}$, 
W.~Przygoda$^{3}$, B.~Ramstein$^{16}$, A.~Reshetin$^{13}$, A.~Rustamov$^{8}$, A.~Sadovsky$^{13}$, 
P.~Salabura$^{3}$, A.~Schmah$^{a}$, E.~Schwab$^{4}$, J.~Siebenson$^{9,10}$, Yu.G.~Sobolev$^{17}$, 
B.~Spruck$^{11}$, H.~Str\"{o}bele$^{8}$, J.~Stroth$^{8,4}$, C.~Sturm$^{4}$, 
A.~Tarantola$^{8}$, K.~Teilab$^{8}$, P.~Tlusty$^{17}$, M.~Traxler$^{4}$, 
H.~Tsertos$^{15}$, T.~~Vasiliev$^{7}$, V.~Wagner$^{17}$, M.~Weber$^{10}$, C.~Wendisch$^{6,d}$, 
J.~W\"{u}stenfeld$^{6}$, S.~Yurevich$^{4}$, Y.~Zanevsky$^{7}$}

\affiliation{
(HADES collaboration) \\\mbox{$^{1}$Istituto Nazionale di Fisica Nucleare - Laboratori Nazionali del Sud, 95125~Catania, Italy}\\
\mbox{$^{2}$LIP-Laborat\'{o}rio de Instrumenta\c{c}\~{a}o e F\'{\i}sica Experimental de Part\'{\i}culas , 3004-516~Coimbra, Portugal}\\
\mbox{$^{3}$Smoluchowski Institute of Physics, Jagiellonian University of Cracow, 30-059~Krak\'{o}w, Poland}\\
\mbox{$^{4}$GSI Helmholtzzentrum f\"{u}r Schwerionenforschung GmbH, 64291~Darmstadt, Germany}\\
\mbox{$^{5}$Technische Universit\"{a}t Darmstadt, 64289~Darmstadt, Germany}\\
\mbox{$^{6}$Institut f\"{u}r Strahlenphysik, Helmholtz-Zentrum Dresden-Rossendorf, 01314~Dresden, Germany}\\
\mbox{$^{7}$Joint Institute of Nuclear Research, 141980~Dubna, Russia}\\
\mbox{$^{8}$Institut f\"{u}r Kernphysik, Goethe-Universit\"{a}t, 60438 ~Frankfurt, Germany}\\
\mbox{$^{9}$Excellence Cluster 'Origin and Structure of the Universe' , 85748~Garching, Germany}\\
\mbox{$^{10}$Physik Department E12, Technische Universit\"{a}t M\"{u}nchen, 85748~Garching, Germany}\\
\mbox{$^{11}$II.Physikalisches Institut, Justus Liebig Universit\"{a}t Giessen, 35392~Giessen, Germany}\\
\mbox{$^{12}$Istituto Nazionale di Fisica Nucleare, Sezione di Milano, 20133~Milano, Italy}\\
\mbox{$^{13}$Institute for Nuclear Research, Russian Academy of Science, 117312~Moscow, Russia}\\
\mbox{$^{14}$Institute of Theoretical and Experimental Physics, 117218~Moscow, Russia}\\
\mbox{$^{15}$Department of Physics, University of Cyprus, 1678~Nicosia, Cyprus}\\
\mbox{$^{16}$Institut de Physique Nucl\'{e}aire (UMR 8608), CNRS/IN2P3 - Universit\'{e} Paris Sud, F-91406~Orsay Cedex, France}\\
\mbox{$^{17}$Nuclear Physics Institute, Academy of Sciences of Czech Republic, 25068~Rez, Czech Republic}\\
\mbox{$^{18}$LabCAF. F. F\'{\i}sica, Univ. de Santiago de Compostela, 15706~Santiago de Compostela, Spain}\\ 
\\
\mbox{$^{a}$ also at Lawrence Berkeley National Laboratory, ~Berkeley, USA}\\
\mbox{$^{b}$ also at ISEC Coimbra, ~Coimbra, Portugal}\\
\mbox{$^{c}$ also at ExtreMe Matter Institute EMMI, 64291~Darmstadt, Germany}\\
\mbox{$^{d}$ also at Technische Universit\"{a}t Dresden, 01062~Dresden, Germany}\\
\mbox{$^{e}$ also at Dipartimento di Fisica, Universit\`{a} di Milano, 20133~Milano, Italy}\\
\mbox{$^{f}$ also at Frederick University, 1036~Nicosia, Cyprus}\\
\mbox{$^{\ast}$ corresponding authors: jia-chii.chen@tum.de, kirill.lapidus@ph.tum.de}
}

\date{\today}

\begin{abstract}
An exclusive analysis of the 4-body final states $\mathrm{\Lambda + p + \pi^{+}  + K^{0}}$ and $\mathrm{\Sigma^{0} + p + \pi^{+} + K^{0}}$ measured with HADES for p+p collisions at a beam kinetic energy of 3.5 GeV is presented. The analysis uses various phase space variables, such as missing mass and invariant mass distributions, in the four particle event selection (p, $\pi^+$, $\pi^+$, $\pi^-$) to find cross sections of the different production channels, contributions of the intermediate resonances $\mathrm{\Delta^{++}}$ and $\mathrm{\Sigma(1385)^{+}}$ and corresponding angular distributions. A dominant resonant production is seen, where the reaction $\mathrm{\Lambda + \Delta^{++} + K^{0}}$ has an about ten times higher cross section ($\mathrm{29.45\pm0.08^{+1.67}_{-1.46}\pm2.06\,\mu b}$) than the analogous non-resonant reaction ($\mathrm{2.57\pm0.02^{+0.21}_{-1.98}\pm0.18\,\mu b}$). A similar result is obtained in the corresponding $\Sigma^{0}$ channels with $\mathrm{9.26\pm0.05^{+1.41}_{-0.31}\pm0.65\,\
mu b}$ in the resonant and $\mathrm{1.35\pm0.02^{+0.10}_{-1.35}\pm0.09\,\mu b}$ in the non-resonant reactions.
\end{abstract}


\maketitle

\section{Introduction}\label{sec:pp_Intro}







The lightest strange particles, kaons, are object of many investigations within modern nuclear physics. Due to strangeness conservation in strong interactions, kaons and antikaons are often produced together in hadron-hadron collisions but interact with nucleons very differently. 
The presence of baryon resonances just below the $\bar{K}N$ threshold makes the antikaon-nucleon scattering very complex, with the necessity of a coupled-channel treatment. The kaon-nucleon interaction is, on the contrary, featureless, relatively weak, and is reduced at low energies to the non-resonant elastic scattering. 

In p+A or A+A collisions there are several mechanisms that can lead to kaon production. 
Kaons can not only be directly produced in nucleon-nucleon collisions but also in multi-step processes through N$\Delta$ or $\Delta\Delta$ reactions (here and further in the text the $\Delta(1232)$-resonance is discussed, unless explicitly stated differently). However, experimentally one can only study the direct kaon production through elementary p+p/n or $\mathrm{\pi}$+p/n collisions. The measurement of exclusive kaon production channels, in particular including intermediate resonances, plays a fundamental role in elementary and heavy ion collisions at kinetic energies in the $\mathrm{GeV}$ regime \cite{Agakishiev:2011qw,Fabbietti:2013npl,Agakishiev:2012qx} and hence these contributions should be evaluated quantitatively.  

The production of strangeness in 3-body final states ($\mathrm{p+p \to p+Y+K}$), where Y stands for a $\Sigma$- or a $\Lambda$-hyperon, was studied in detail for a wide range of energies up to $2$~GeV both from the experimental
and the theoretical
sides \cite{Balestra:1999br,Maggiora:2001tx,AbdelBary:2010pc, AbdelBary:2012vw, AbdelBary:2012zz}. Inclusive cross-sections of the 3-body final states including $\Sigma$- and 
$\Lambda$-hyperons have been measured accurately but so far no conclusive picture was drawn that was able to explain the role of intermediate N$^*$ resonances decaying into $\Lambda$-kaon 
pairs to the final states. In particular, at the same beam energies studied in this paper, the measured $\mathrm{p+\Lambda+K}$ final state is found to be difficult to interpret even including several N$^*$'s with
masses ranging from $1700$ to $1900$~MeV \cite{Fabbietti:2013npl}. There and probably also at lower energies partial wave techniques have to be employed \cite{Ceci:2013uoa} to disentangle the different
contributions of the resonant and non-resonant terms.

On the other hand, 4-body final states, with the simplest final state $\mathrm{p+p \to Y+p+\pi+K}$ having a threshold of $\sqrt{s} \approx 2.68$~GeV, did not get enough attention so far. The region 
corresponding to $\sqrt{s} > 3.35$~GeV was explored by a number of bubble chamber experiments \cite{Bierman:1966zz, Alexander:1967zz, Klein:1970ri, Firebaugh:1968rq}, but for the 
$\Sigma$-associated channels $\mathrm{p+p \to \Sigma+p+\pi+K}$ no experimental data exist below $\sqrt{s} = 3.35$~GeV. For the $\Lambda$-associated channels the situation is the same, with an
 exception of a $\mathrm{p+p \to \Lambda+p+\pi^{+}+K^{0}}$ measurement done at $\sqrt{s} = 2.97$~GeV \cite{Nekipelov:2006if}.
 
 The state of theoretical calculations for strangeness production in the 4-body final state is even worse: only one model \cite{Tsushima:1998jz} gives predictions for the energy dependence
  of the intermediate reactions $\mathrm{N+N \to \Delta+Y+K}$. In this approach an accompanying pion appears exclusively through the decay of a $\Delta$ resonance. The model significantly 
  overestimates existing measurements of the reaction strengths, in particular for the channel $\mathrm{p+p \to \Lambda+p+\pi^{+}+K^{0}}$.

The assumption that the $\Delta$ resonance contribution to 4-body reactions is dominant should be verified experimentally. For this purpose, we show in this work the analysis of the 
channel $\mathrm{p+p \rightarrow \Lambda/\Sigma^{0} + p + \pi^{+} + K^{0}}$  and the extracted contribution from the formation of an intermediate $\Delta^{++}$ resonance
 \cite{Nekipelov:2006if, Klein:1970ri, Firebaugh:1968rq} to this final state. The data presented here have been measured with the HADES \cite{Agakishiev:2009am} detector setup at GSI (Darmstadt).
An analogous study has never been performed for the reaction $\mathrm{p+p \rightarrow \Sigma^0 + p + \pi^{+} + K^{0}}$, in which the proton and the $\pi^+$ might stem from the $\Delta^{++}$ decay, probably because the separation of $\Lambda$ and $\Sigma^0$ requires an excellent invariant and missing mass resolution. 
Not only because of the good resolution, but also because of the large acceptance, HADES is an excellent detector to investigate the mentioned reactions. 

Besides the analysis of the exclusive reaction channels $\mathrm{p+p \to Y+\Delta^{++}/(p\pi^{+})+K^{0}}$ at the beam kinetic energy of $3.5$~GeV corresponding to $\sqrt{s} = 3.18$~GeV, the channels $\mathrm{p+p \rightarrow \Sigma^+ + p + K^{0}}$ and $\mathrm{p+p \rightarrow \Sigma(1385)^+ + p + K^{0}}$ have been investigated and results are also reported in this work. These measurements are of great interest for the inclusive analysis of $\mathrm{K^0}$ production in p+p \cite{Berger-Chen:2012wxa} and p+Nb collisions measured at the same beam energy \cite{Agakishiev:2013zzz}. Moreover, they complement the exclusive analysis of the 3-body $\mathrm{p+K^{+}+\Lambda}$ final state aimed at the search for the kaonic bound state \cite{Fabbietti:2013npl}.


These measurements probe an interesting energy region that is challenging for theoretical descriptions. At lower energies, close to the threshold of strangeness production, where the number of opened channels is small and restricted mostly to 3-body reactions, the so-called resonance models are widely applied. In this approach all hyperon-kaon pairs originate from the decay of an intermediate N$^{*}$ or $\Delta^{*}$ baryonic resonance formed in two-body processes: $\mathrm{N+N \to N+R \to N+Y+K}$. 
At high energies multi-particle production is usually treated with the help of string fragmentation models such as {\sc pythia}. The energy explored in the present study lies in the transition region between resonance and string fragmentation models. Note that for the description of electron-positron pair production performed by the HADES collaboration in the same data set both the {\sc pythia} and the resonance model approach have been applied \cite{HADES:2011ab, Weil:2012ji}. Precise measurements of further hadron production channels, including strangeness, are, therefore, necessary to constrain any theoretical calculations or models.

While p+A reactions are dominated by NN collisions, in heavy-ion collisions processes including a $\Delta$-resonance become more important especially at energies below the kaon production threshold. At high energies NN collisions start to dominate the production mechanism again \cite{Hartnack:2011cn}. 

Due to the complex dynamics of collisions involving heavy ions, the interpretation of experimental results is usually assisted by the comparison to the outcome of transport models that aim to incorporate all available knowledge about production and in-medium propagation of hadrons. 
Clearly, strengths of strangeness production in underlying nucleon-nucleon reactions constitute a major ingredient of this knowledge \cite{Fuchs:2005zg}. 
However, for the correct treatment of strangeness production not only energy dependent cross sections of different channels are required, but also more detailed information about the different production mechanisms which lead to the same final state. Hence, also information such as angular distributions of final state particles and contributions by intermediate resonance states as presented in this work are extremely important.

This paper is structured as follows. Section~\ref{sec:pp_HADESExp} describes shortly the experimental apparatus and section~\ref{sec:pp_DataAna} contains the analysis of the exclusive channels and the comparison to simulated data with the extraction of the angular distribution and cross-section values. Section~\ref{sec:pp_SumCon} is dedicated to the summary and discussion of the presented results.


\section{The Experiment}\label{sec:pp_HADESExp}
The \textbf{H}igh-\textbf{A}cceptance \textbf{D}i-\textbf{E}lectron \textbf{S}pectrometer (HADES) is a versatile experiment currently operating at the SIS18 synchrotron (GSI Helmholtzzentrum in Darmstadt, Germany) delivering kinetic beam energies in the range of $1-2\,\mathrm{AGeV}$ for nucleus-nucleus collisions. For proton induced reactions, energies up to $3.5\,\mathrm{GeV}$ can be reached. The detector system has a 85$\%$ azimuthal coverage, while the polar angles are covered from $18^{\circ}$ to $85^{\circ}$ degrees. The momentum resolution is $\mathrm{\Delta}$ p/p $\approx$ 3$\%$.

The multi wire drift chambers (MDCs) - two planes in front of and two behind the superconducting magnet (toroidal field) employed for charged particle tracking - and the Time-Of-Flight wall are the detector components which have been used for the analysis presented here. The MDCs deliver the momentum information and the particle identification via the specific energy loss dE/dx for each particle, the Time-Of-Flight wall has been used to set the online trigger conditions. In the present experiment the first level trigger (LVL1) condition required at least three hits in the Time-Of-Flight wall (M3) to reduce contributions from the p+p elastic scattering. For more details see \cite{Agakishiev:2009am}.

In this analysis proton-proton reactions at a kinetic beam energy of $\mathrm{E_{kin}} = \,3.5\,\mathrm{GeV}$ have been studied. The beam had an average intensity of $\sim\,1\times$10$^{7}$ particles/s. As a target material liquid hydrogen with a density of $0.35\,\mathrm{g/cm^2}$ has been used. The dimensions of the cylindric target cell were $15\,\mathrm{mm}$ in diameter with a length of $44\,\mathrm{mm}$ \cite{Rustamov:2010zz}. The corrsponding total interaction probability amounts to $0.7\,\%$. In total, $1.14\times10^9$ events were collected with this setup.

\begin{figure}[htb]
\includegraphics[width=0.45\textwidth]{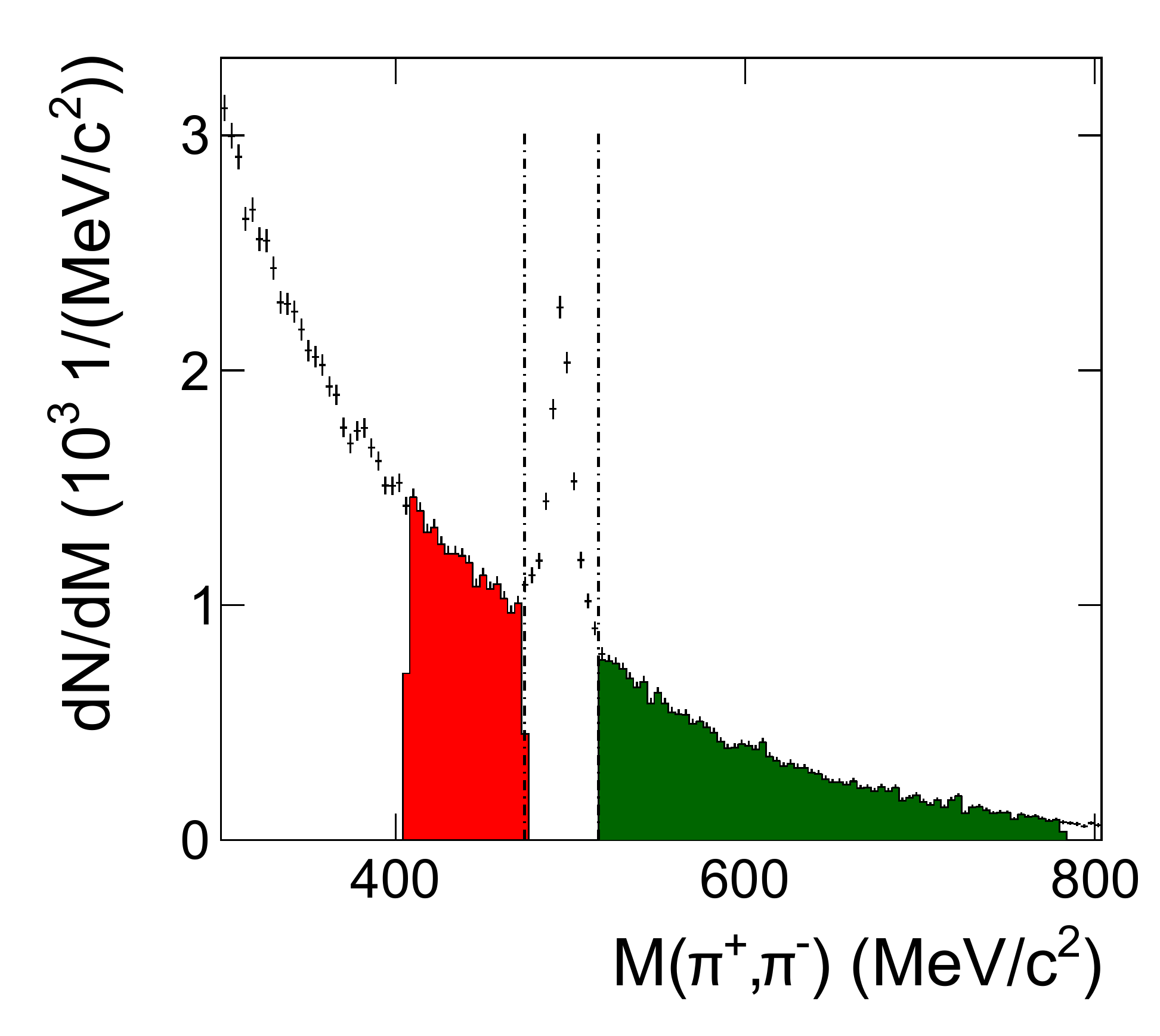}
\caption{\label{fig:pp_ExclInvMassK0s} Color online. $\mathrm{\pi^+\pi^-}$-invariant mass distribution after secondary vertex cuts for the selected event sample. The reconstructed values from a fit with the sum of two Gaussians, a polynomial and a Landau function are $\mathrm{<m_{K^0_S}> = 494.9\,MeV/c^2}$, $\mathrm{<\sigma_{K^0_S}> = 7.1\,MeV/c^2}$, $\mathrm{N_{K^0_S} = 6102}$ and S/B 0.64. The dashed-dotted lines show the 3$\mathrm{\sigma}$ region around the $\mathrm{K^0_S}$ signal. The red and green areas indicate the chosen low and high mass sideband sample respectively.}
\end{figure}
\section{Data Analysis}\label{sec:pp_DataAna}
One of the uncertainties in the transport models is the cross section of $\mathrm{K^{0}}$ production channels in nucleon-nucleon collisions associated with the production of $\mathrm{\Delta(1232)}$-resonances. In this section, we show the data analysis procedure to determine the rate of $\mathrm{K^0_S}$ production associated with a hyperon and a nucleon (or a $\mathrm{\Delta^{++}}$) in events containing a proton, a $\mathrm{\pi^+}$, another $\mathrm{\pi^+}$ and a $\mathrm{\pi^-}$. To this end, $\mathrm{K^0_S}$ candidates have been reconstructed by the invariant mass of the $\mathrm{\pi^+\pi^-}$-pairs, leading also to combinatorial background stemming from non-strange production channels. In order to understand and later on to subtract the background which survives the $\mathrm{K^0_S}$ selection cuts, a sideband event sample has been defined. This method is described in section~\ref{sec:pp_SidebandAna}. Further contributions by various $\mathrm{K^0}$ reaction channels have been simulated with the Pluto~\cite{
Frohlich:2007bi} event generator, filtered through a full scale simulation and analyzed the same way as real events. A simultaneous fit of all these channels to five kinematic observables (three missing mass and two invariant mass distributions) has been carried out to extract the cross sections associated to each production channel.  For the reactions $\mathrm{p+p \rightarrow \Lambda + \Delta^{++} + K^{0}}$ and $\mathrm{p+p \rightarrow \Sigma^{0} + \Delta^{++} + K^{0}}$ also the angular distributions corrected for acceptance and efficiency have been determined.

\subsection{Event and Track Selection}\label{sec:pp_EvSelect}
As mentioned above, we have focused on the two reactions $\mathrm{p+p \rightarrow \Lambda + \Delta^{++} + K^{0}}$ and $\mathrm{p+p \rightarrow \Sigma^{0} + \Delta^{++} + K^{0}}$ and developed an exclusive event selection, which we can distinguish from each other by reconstructing first the $\mathrm{K^{0}_S}$. Therefore, only events with exactly the four charged particles proton, $\mathrm{\pi^+}$, $\mathrm{\pi^+}$, $\mathrm{\pi^-}$ have been selected assuming that the $\mathrm{\pi^-}$ and one of the $\mathrm{\pi^+}$ stem from a $\mathrm{K^0_S}$ decay (BR = $69.2 \,\%$), whereas the proton and the other $\mathrm{\pi^+}$ are either produced directly or originate from the $\mathrm{\Delta^{++}}$ decay. Due to combinatorics, it is possible to use one event multiple times even after applying off-target, off-vertex and a $\mathrm{K^0_S}$ mass cut (see next paragraph for a detailed description). However, by rejecting events with more than four detected particles the probability to use an event more than one time has 
been decreased from $23.3\,\%$ to less than $0.3\,\%$.

To reduce the contribution from off-target events a three-dimensional cut is applied on the primary vertex position. For each event the primary vertex has been calculated as the averaged intersections of the reconstructed $\mathrm{K^0_S}$ track and the remaining proton and $\mathrm{\pi^+}$ tracks. A $17\,\mathrm{mm}$ wide interval in the XY-plane around the nominal beam position and a $60\,\mathrm{mm}$ interval in the Z direction have been used to define the target region. The identification of the particles has been carried out by two-dimensional cuts on the dE/dx vs. polarity$\times$momentum distribution, in which the positive pions are well separated from the protons up to a momentum of $1\,\mathrm{GeV/c}$. As mentioned above $\mathrm{K^0_S}$ are reconstructed from their decay particles $\mathrm{\pi^+\pi^-}$. Therefore, all possible combinations of $\mathrm{\pi^+\pi^-}$-pairs have been formed per event and their secondary vertices have been calculated. An effective suppression of the combinatorial 
background is achieved by applying the following secondary vertex cuts: (1) distance between the two pion tracks ($\mathrm{d_{\pi^+-\pi^-}< 7\,mm}$), (2) distance between the primary reaction and the secondary decay vertex ($\mathrm{d(K^0_S - V)\,> 25\,mm}$), (3) distances of closest approach of the two pion tracks with respect to the primary vertex ($\mathrm{DCA_{\pi^+}> \,7\,mm}$, $\mathrm{DCA_{\pi^-}> 7\,mm}$). The resulting $\mathrm{\pi^+\pi^-}$-invariant mass spectrum is depicted in Fig.~\ref{fig:pp_ExclInvMassK0s}. It shows a clear peak over the remaining combinatorial background, which corresponds to the $\mathrm{K^0_S}$ signal. This invariant mass spectrum is fitted with the sum of two Gaussians for the signal and a polynomial and a Landau function for the description of the background. Since only final states containing a $\mathrm{K^0}$ have been considered in this analysis, a 3$\mathrm{\sigma}$ cut around the nominal $\mathrm{K^0_S}$ mass has been applied, where $\mathrm{\sigma}$ is the averaged 
standard deviation of the two Gaussians. The cut boundaries are indicated by the dashed-dotted lines in Fig.~\ref{fig:pp_ExclInvMassK0s}. 

\begin{figure}[h]
\includegraphics[width=0.45\textwidth]{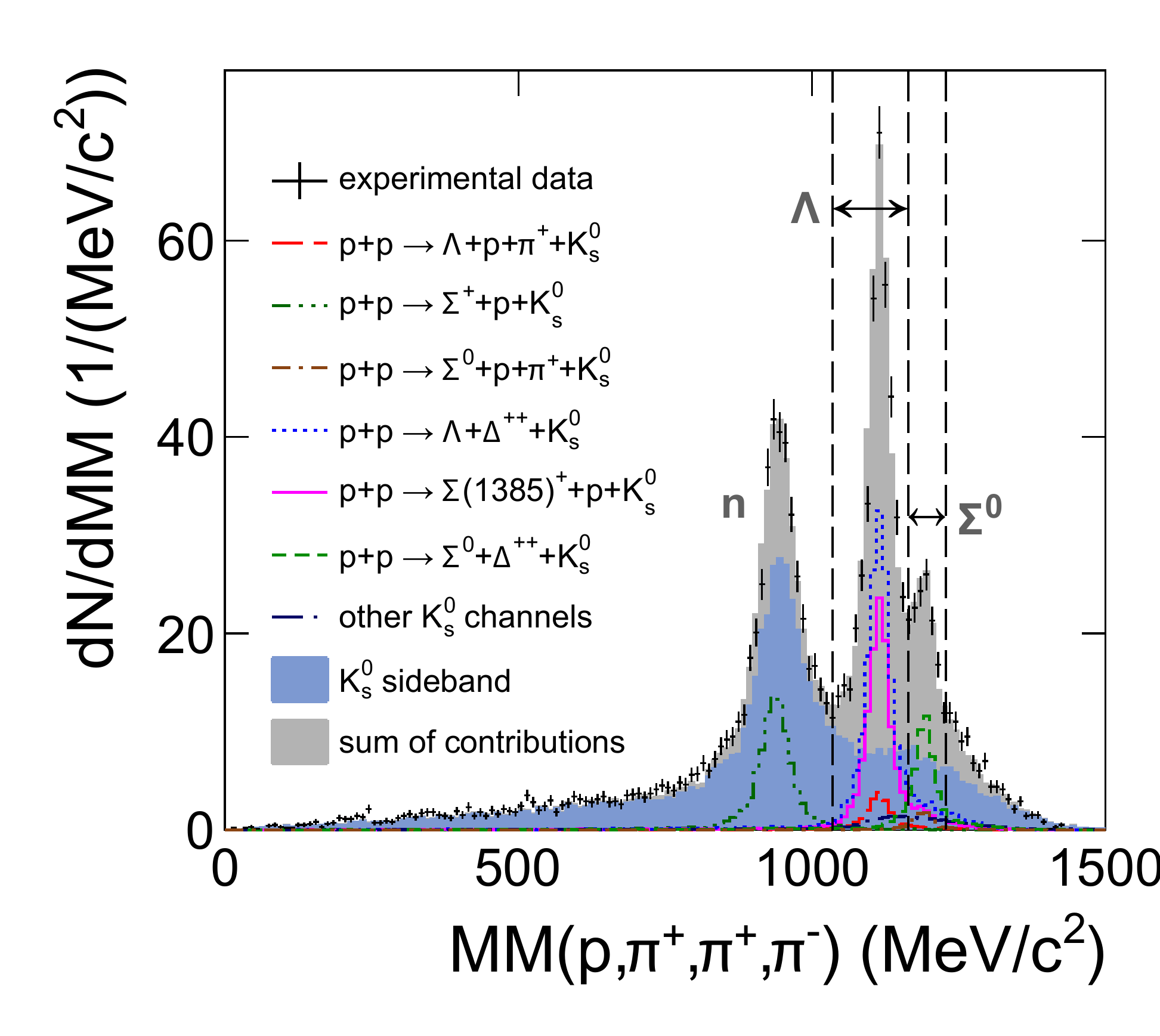}
\caption{\label{fig:pp_ExclMMABCD} Color online. Missing mass distribution with respect to the p, $\pi^+$, $\pi^+$ and $\pi^-$ with a cut on the $\mathrm{K^0_S}$ mass in the $\pi^+\pi^-$-invariant mass spectrum (Fig.~\ref{fig:pp_ExclInvMassK0s}). The gray histogram corresponds to the sum of simulated contributions plus the background defined by the sideband sample. The double arrows indicate the so-called $\mathrm{\Lambda}$-cut and $\mathrm{\Sigma^0}$-cut, respectively. See text for details.}
\end{figure}

After this $\mathrm{K^0_S}$ preselection the missing mass to proton, $\mathrm{\pi^+}$, $\mathrm{\pi^+}$ and  $\mathrm{\pi^-}$ (MM(p,$\pi^+$,$\pi^+$,$\pi^-$)) can be studied. Fig.~\ref{fig:pp_ExclMMABCD} shows the obtained $\mathrm{MM(p,\pi^+,\pi^+,\pi^-)}$ distribution, in which one can clearly identify the $\mathrm{\Lambda}$ and the $\mathrm{\Sigma^{0}}$ peaks associated to the reactions $\mathrm{p+p \rightarrow \Lambda/\Sigma^0 + p +\pi^+ + K^{0}}$ and $\mathrm{p+p \rightarrow \Sigma(1385)^{+} + p + K^{0}}$. The peak at the neutron mass is due to the reactions without strangeness production and a contribution from the channel $\mathrm{p+p \rightarrow \Sigma^{+} + p + K^{0}}$.
    
\begin{figure}[h]
\includegraphics[width=0.45\textwidth]{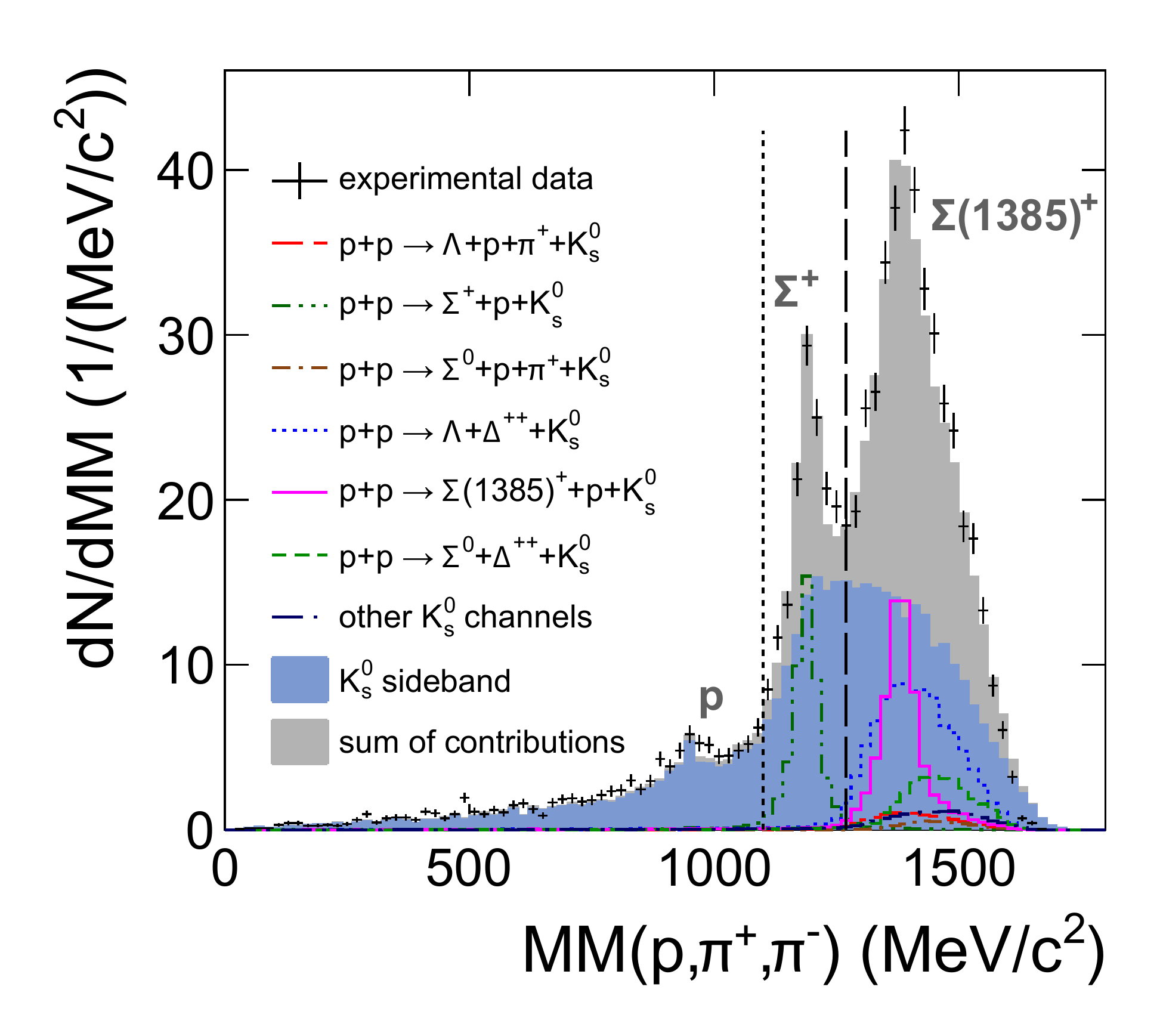}
\caption{\label{fig:pp_ExclMMABD} Color online. Missing mass distribution with respect to the p, $\pi^+$ and $\pi^-$ with a cut on the $\mathrm{K^0_S}$ mass in the $\pi^+\pi^-$-invariant mass spectrum (Fig.~\ref{fig:pp_ExclInvMassK0s}). The gray histogram corresponds to the sum of simulated contributions plus the background defined by the sideband sample. The dotted and dashed lines at $1100\,\mathrm{MeV/c^2}$ and $1270\,\mathrm{MeV/c^2}$ are used as cuts in other variables.}
\end{figure}

Also the missing mass to the proton, $\mathrm{\pi^+}$ and $\mathrm{\pi^-}$ (MM(p,$\pi^+$,$\pi^-$)) has been studied (Fig.~\ref{fig:pp_ExclMMABD}). As will be discussed later in this work, the MM(p,$\pi^+$,$\pi^-$) is employed to extract the contribution of those $\mathrm{K^0}$ production channels associated with a $\mathrm{\Sigma^{+}}$ or a $\mathrm{\Sigma(1385)^{+}}$. The visible proton peak in the distribution corresponds to the remaining non-strange channel $\mathrm{p+p \rightarrow p + n + \pi^+ + \pi^+ + \pi^-}$. Such reactions are also responsible for the low mass tails in the missing mass spectra.
%

\subsection{Sideband Analysis}\label{sec:pp_SidebandAna}
\begin{table}[htb]
\begin{center}
\caption{\label{tab:pp_NonStrange} Possible non-strange reactions contributing to the final state selected in this analysis.}   
\begin{ruledtabular}
\begin{tabular}{l}
Non-strange reactions in selected events\\
\colrule
$p+p \rightarrow p+n+\pi^++\pi^++\pi^-$\\
$p+p \rightarrow p+p+\pi^++\pi^++\pi^-+\pi^-$\\
$p+p \rightarrow p+p+\pi^++\pi^-$\\
$p+p \rightarrow p+p+\eta$\\
\end{tabular}
\end{ruledtabular}
\end{center}
\end{table}
\begin{figure}[h]
\includegraphics[width=0.45\textwidth]{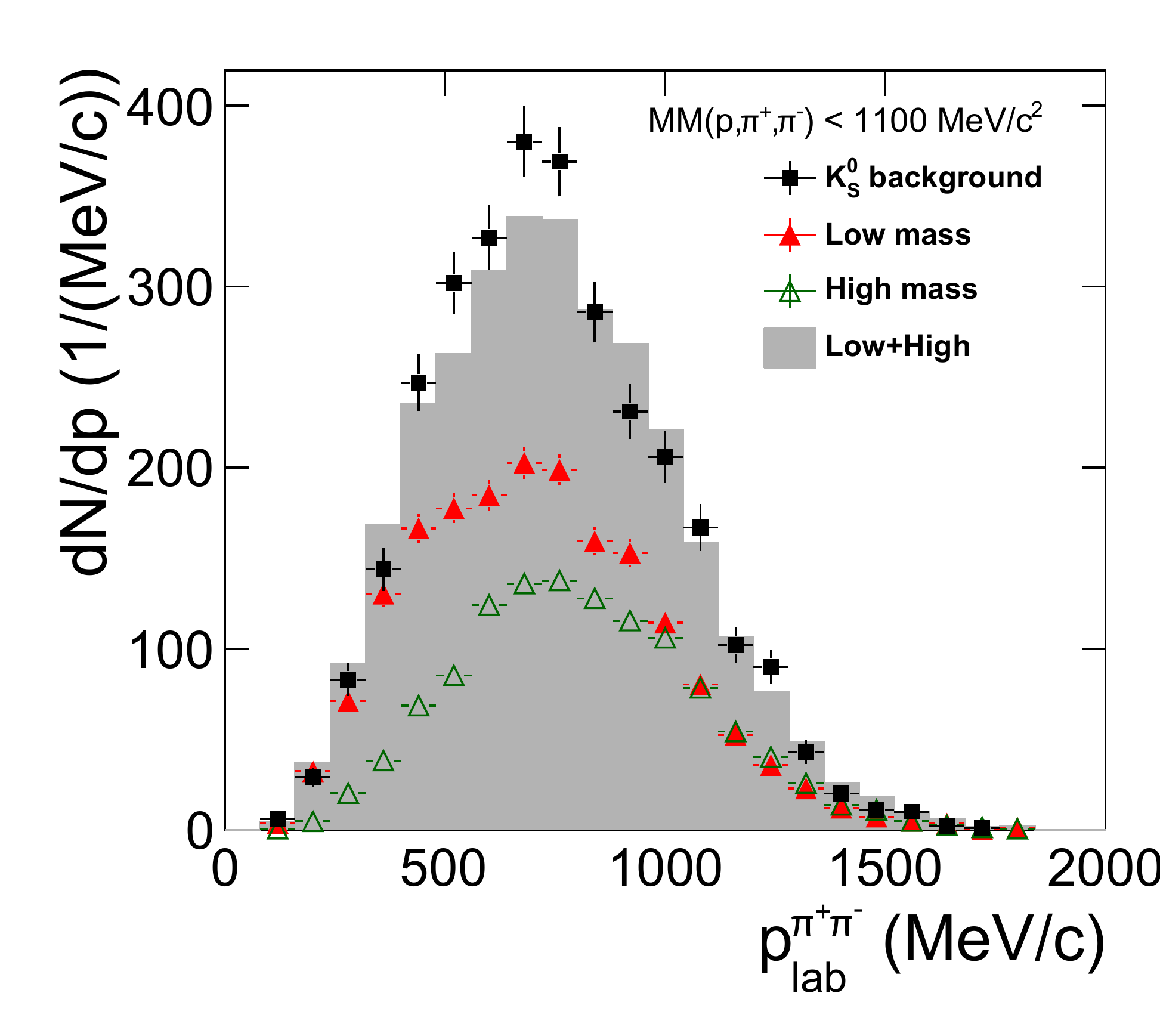}
\caption{\label{fig:pp_ExclMomSideband} Color online. Momentum distribution of the $\mathrm{K^0_S}$ background (full squares). Applied conditions are a 3$\sigma$ cut around the $\mathrm{K^0_S}$ peak in the $\pi^+\pi^-$-invariant mass distribution (Fig.~\ref{fig:pp_ExclInvMassK0s}), and a cut on MM(p,$\pi^+,\pi^-$) $<$ 1100~MeV/c$^2$. Relative weights of low mass (full triangles) and high mass (open triangles) sideband samples were fitted simultaneously to the spectrum. The sum of both is shown in magenta as a gray filled histogram.}
\end{figure}
\begin{figure*}[t]
\includegraphics[width=0.7\textwidth]{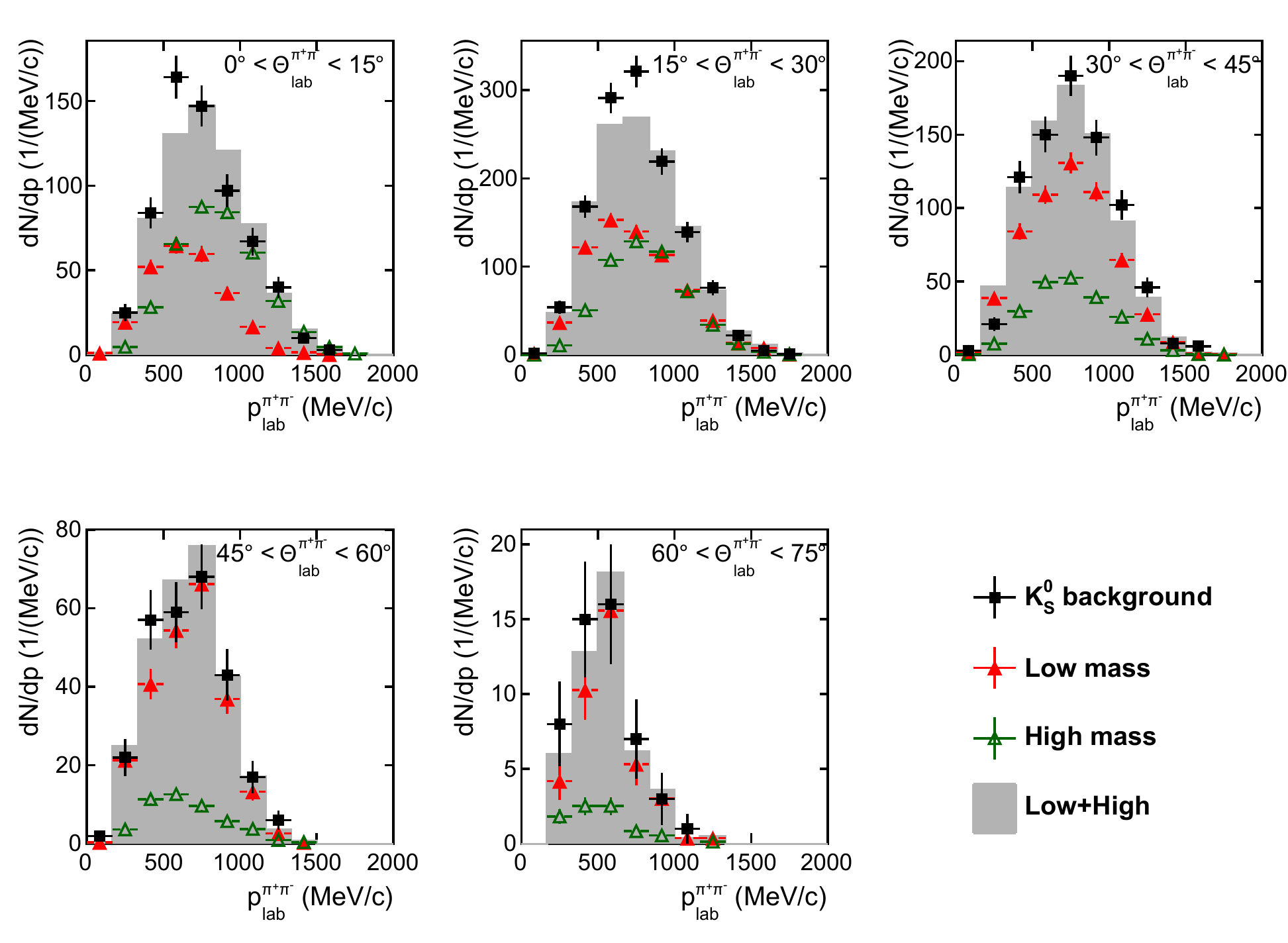}
\caption{\label{fig:pp_ExclPThSideband} Color online. Momentum spectra for several bins in $\mathrm{\Theta_{lab}}$ of the $\mathrm{K^0_S}$ background. Applied conditions are a 3$\sigma$ cut around the $\mathrm{K^0_S}$ peak in the $\pi^+\pi^-$-invariant mass distribution (Fig.~\ref{fig:pp_ExclInvMassK0s}), and a cut on MM(p,$\pi^+,\pi^-$) $<$ 1100~MeV/c$^2$. The scaling of the low and high mass sideband is a result of the simultaneous fit of both sidebands to the momentum distribution of the $\mathrm{K^0_S}$ background in Fig.~\ref{fig:pp_ExclMomSideband}. The symbols are the same as in Fig.~\ref{fig:pp_ExclMomSideband}.}
\end{figure*}
As mentioned above, the selected data sample includes combinatorial background from the $\mathrm{K^0_S}$ reconstruction. But, it also contains contributions from reactions with a non-strange final state. These reactions constitute the main sources of background in the study presented here. Some examples of the background-reactions are listed in Table~\ref{tab:pp_NonStrange}. To evaluate the background contribution to the MM(p,$\pi^+,\pi^+,\pi^-$) and MM(p,$\pi^+,\pi^-$) distributions (Figs.~\ref{fig:pp_ExclMMABCD} and \ref{fig:pp_ExclMMABD}), a sideband analysis has been performed. The sideband sample is defined by selecting regions of the $\pi^+\pi^-$-invariant mass spectrum adjacent to the $\mathrm{K^0_S}$ peak. The selection applied is shown in Fig.~\ref{fig:pp_ExclInvMassK0s}, where the sideband intervals are indicated by the red (low mass, LM) and green (high mass, HM) filled regions. These two samples include negligible contributions from the $\mathrm{K^0_S}$ signal. Both the LM and HM intervals are 
chosen such to have nearly the same integral. The data sample selected by this method should be equivalent in terms of kinematic distributions to the background below the $\mathrm{K^0_S}$ peak in the $\mathrm{\pi^+\pi^-}$-invariant mass that we want to emulate. As it is visible in the missing mass spectrum of Fig.~\ref{fig:pp_ExclMMABD}, the comparison of the experimental distribution to the incoherent sum of the different simulated channels shows that no final state including a reconstructed $\mathrm{K^0_S}$ contributes to the region where MM(p,$\pi^+,\pi^-) <\, 1100\,\mathrm{MeV/c^2}$ (vertical dotted line in Fig.~\ref{fig:pp_ExclMMABD}). Hence, by selecting events which fulfill the condition MM(p,$\pi^+,\pi^-) <\, 1100\,\mathrm{MeV/c^2}$, one obtains a rather pure background sample which can be used as a reference to cross check the kinematic variables of the sideband samples (LM and HM) selected on the  $\mathrm{\pi^+\pi^-}$-invariant mass spectrum. Fig.~\ref{fig:pp_ExclMomSideband} shows the momentum 
distribution of the $\mathrm{\pi^+\pi^-}$-pairs from the pure background events extracted by the missing mass selection (full squares) together with the distribution for the LM and HM samples (full and empty triangles respectively). The distributions from the LM and HM samples have been fitted simultaneously such that their sum (gray area) reproduces the momentum spectrum of the pure background events. This fit results  in a $\mathrm{\chi^2/NDF}$ = 2.32. Fig.~\ref{fig:pp_ExclPThSideband} shows the comparison of the momentum distributions of the different background samples for different bins of the laboratory polar angle $\mathrm{\Theta_{lab}}$. Obviously, the LM and HM samples describe the kinematics of the $\mathrm{K^0_S}$ background properly, for example double differentially in the momentum and $\mathrm{\Theta_{lab}}$ variables. Thus, the common LM+HM sideband sample can be used to describe the background in the whole kinematic range of the selected data.

\subsection{Simulation of the $\mathrm{K^0}$ production channels and simultaneous fit of the experimental data}\label{sec:pp_SimulationFit}
By looking at Fig.~\ref{fig:pp_ExclMMABCD}, one can clearly recognize the peaks corresponding to the $\mathrm{\Lambda}$ and the $\mathrm{\Sigma^0}$ mass, while in Fig.~\ref{fig:pp_ExclMMABD} signals from $\mathrm{\Sigma^+}$ and $\mathrm{\Sigma(1385)^+}$ show up. However, the missing mass distributions do not only contain background events, that can be described by the sideband method, but also several other contributions from $\mathrm{K^0}$ production channels that overlap. In order to evaluate these different contributions an incoherent cocktail composed of the $\mathrm{K^0}$ production channels listed in Table~\ref{tab:pp_K0Cocktail_Table} has been simulated using the Pluto event generator \cite{Frohlich:2007bi}. This reaction list has been grouped into three different classes according to their relative abundances in the detected final state and to the region the corresponding missing mass spectra contribute to. The first class (C1) contains the $\mathrm{K^0}$ reactions that survive with rather high 
probability the event selection of the exactly four charged particles (p, $\mathrm{\pi^+}$, $\mathrm{\pi^+}$, $\mathrm{\pi^-}$). The second class (C2) includes reactions that lead to multiple pion production, thus have a lower probability to fulfill the multiplicity selection, but due to the limited geometrical acceptance of HADES can still deliver a contribution. As these channels display similar shapes in the missing mass spectrum, it is legitimate to group them. The same statement holds for the third class (C3) with the remaining $\mathrm{K^0}$ reactions. The contribution of the reactions $\mathrm{p+p \rightarrow Y + p + K^{*}(892)^{+}}$ with Y being a $\mathrm{\Lambda}$ ($\mathrm{\epsilon = 230\,MeV}$) or a $ \mathrm{\Sigma^0}$ ($\mathrm{\epsilon = 157\,MeV}$) have been considered, but no phase-space and acceptance are provided in the selected data sample. All the simulated reactions have been processed using the same full-scale analysis employed for the experimental data, thus taking into account the 
efficiency of the trigger condition (M3), the tracking algorithm and the analysis procedure. The particle decay, the acceptance and the materials of the HADES detector have been considered by using {\sc geant3}. Moreover, special attention had been payed on the mass distribution of the $\mathrm{\Delta^{++}}$ resonance employed in the simulation. Here, a relativistic Breit-Wigner distribution with the parameters m (running unstable mass), $\mathrm{M_R}$ (static pole mass of the resonance) and $\Gamma^{tot}(m)$ (mass dependent width) is implemented in the Pluto event generator. Details can be found in \cite{Frohlich:2007bi}.
\begin{table}[h]
\begin{center}
\caption{\label{tab:pp_K0Cocktail_Table} $\mathrm{K^0}$ production channels contributing to the selected final state.
 The cross sections $\mathrm{\sigma^{fit}_{ch}}$ at $\mathrm{3.5\, GeV}$ are determined by a fit with a cross section parametrization from Eq. 34 in \cite{Sibirtsev:1998dh} to experimental cross sections measured at other energies (*no experimental data existing to perform the fit). The excess energies $\mathrm{\epsilon}$ are calculated for p+p reactions at $\mathrm{3.5\, GeV}$.}
\begin{ruledtabular}
\begin{tabular}{lll}
Main contributing reactions (C1)&$\mathrm{\sigma^{fit}_{ch}\,[\mu b]}$&$\mathrm{\epsilon\,[MeV]}$\\
\colrule
$p+p \rightarrow \Sigma^++p+K^0$&20.43&551\\
$p+p \rightarrow \Lambda+p+\pi^++K^0$&18.40&485\\
$p+p \rightarrow \Sigma^0+p+\pi^++K^0$&12.38&408\\
$p+p \rightarrow \Lambda+\Delta^{++}+K^0$&4.47&331\\
$p+p \rightarrow \Sigma^0+\Delta^{++}+K^0$&-*&254\\
$p+p \rightarrow \Sigma(1385)^++p+K^0$&5.31&358\\
\colrule
Multi-pion $\mathrm{K^0}$ reactions (C2)&$\mathrm{\sigma^{fit}_{ch}\,[\mu b]}$&$\mathrm{\epsilon\,[MeV]}$\\
\colrule
$p+p \rightarrow \Lambda+n+\pi^++\pi^++K^0$&5.08&344\\
$p+p \rightarrow \Lambda+p+\pi^++\pi^0+K^0$&4.46&350\\
$p+p \rightarrow \Sigma^-+p+\pi^++\pi^++K^0$&3.75&264\\
$p+p \rightarrow \Sigma^++p+\pi^++\pi^-+K^0$&2.26&272\\
\colrule
Other $\mathrm{K^0}$ reactions (C3)&$\mathrm{\sigma^{fit}_{ch}\,[\mu b]}$&$\mathrm{\epsilon\,[MeV]}$\\
\colrule
$p+p \rightarrow p+n+K^++\bar{K^0}$&7.58&307\\
$p+p \rightarrow \Sigma^++n+\pi^++K^0$&4.53&410\\
$p+p \rightarrow \Sigma^++p+\pi^0+K^0$&4.06&416\\
$p+p \rightarrow \Sigma^++\Delta^{++}+K^0$&6.59&257\\
$p+p \rightarrow p+p+\pi^++K^-+K^0$&2.02&169\\
\end{tabular}
\end{ruledtabular}   
\end{center}
\end{table}

\begin{figure}[h]
\includegraphics[width=0.45\textwidth]{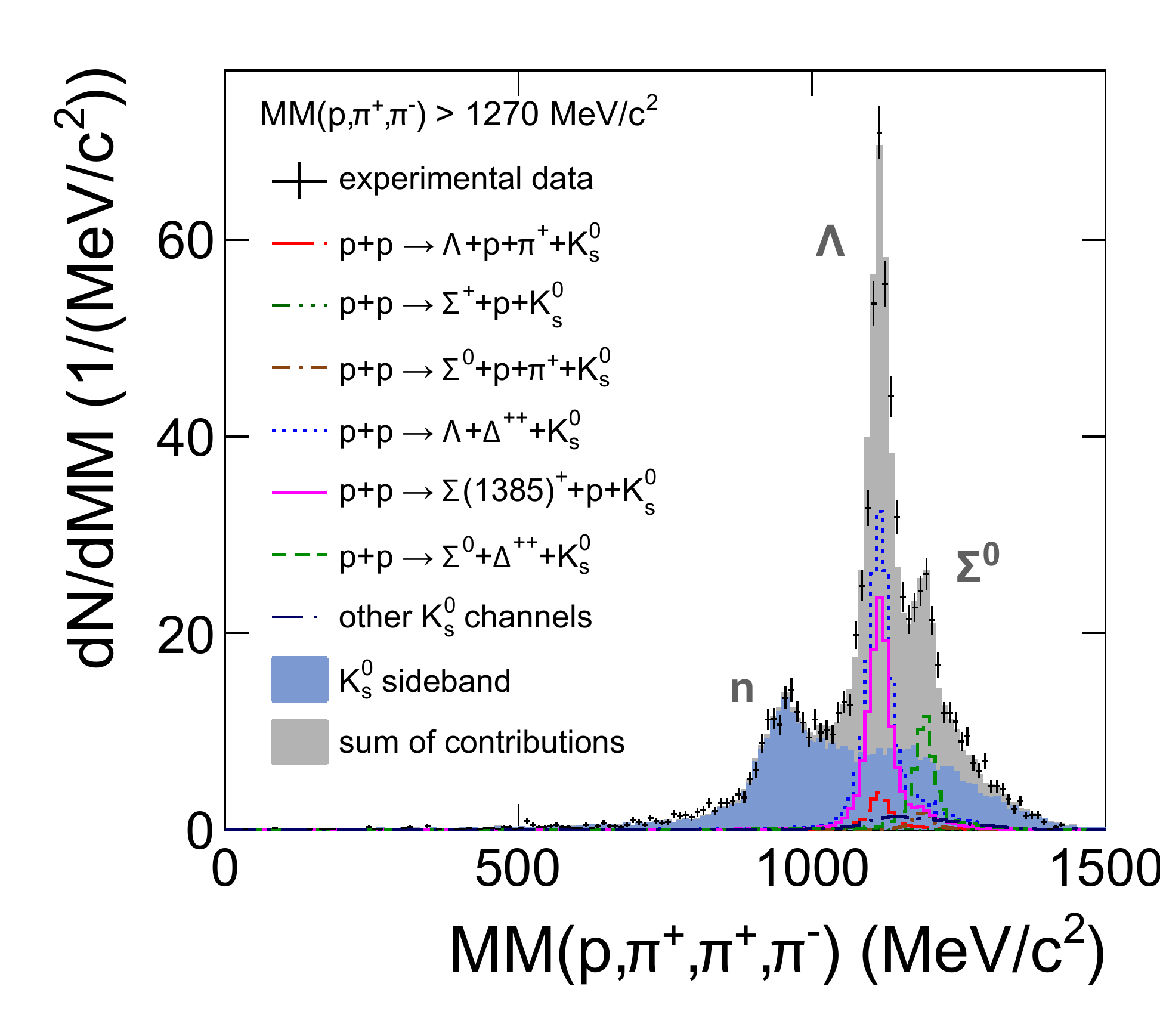}
\caption{\label{fig:pp_ExclMMABCD_hhcutMMABD} Color online. Missing mass distribution with respect to the p, $\pi^+$, $\pi^+$ and $\pi^-$ obtained after the cut MM(p,$\pi^+,\pi^-) >\, 1270\,\mathrm{MeV/c^2}$ and a cut on the $\mathrm{K^0_S}$ mass in the $\pi^+\pi^-$-invariant mass spectrum (Fig.~\ref{fig:pp_ExclInvMassK0s}). The grey histogram corresponds to the sum of simulated contributions plus the background defined by the sideband sample.}
\end{figure}

\begin{figure}[h]
\includegraphics[width=0.45\textwidth]{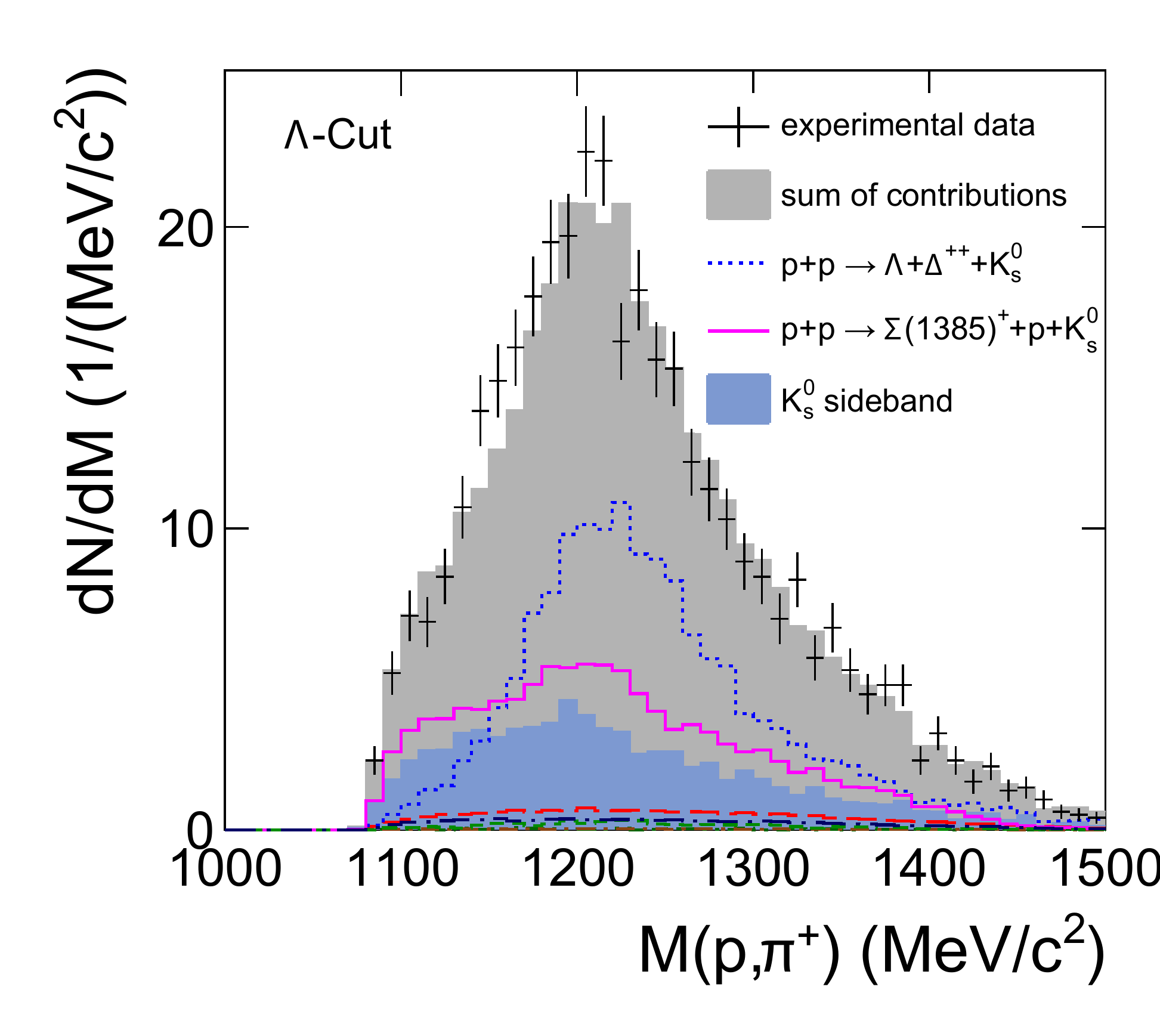}
\caption{\label{fig:pp_ExclIML} Color online. $\mathrm{p\pi^+}$-invariant mass distribution after the $\Lambda$-cut on the MM(p,$\pi^+,\pi^+,\pi^-$) distribution (Fig.~\ref{fig:pp_ExclMMABCD}) and with a cut on the $\mathrm{K^0_S}$ mass in the $\pi^+\pi^-$-invariant mass spectrum (Fig.~\ref{fig:pp_ExclInvMassK0s}). The gray histogram corresponds to the sum of simulated contributions plus the background defined by the sideband sample. The same color code and line styles are used as in Figs.~\ref{fig:pp_ExclMMABCD},~\ref{fig:pp_ExclMMABD} and \ref{fig:pp_ExclMMABCD_hhcutMMABD}.}
\end{figure}

\begin{figure}[h]
\includegraphics[width=0.45\textwidth]{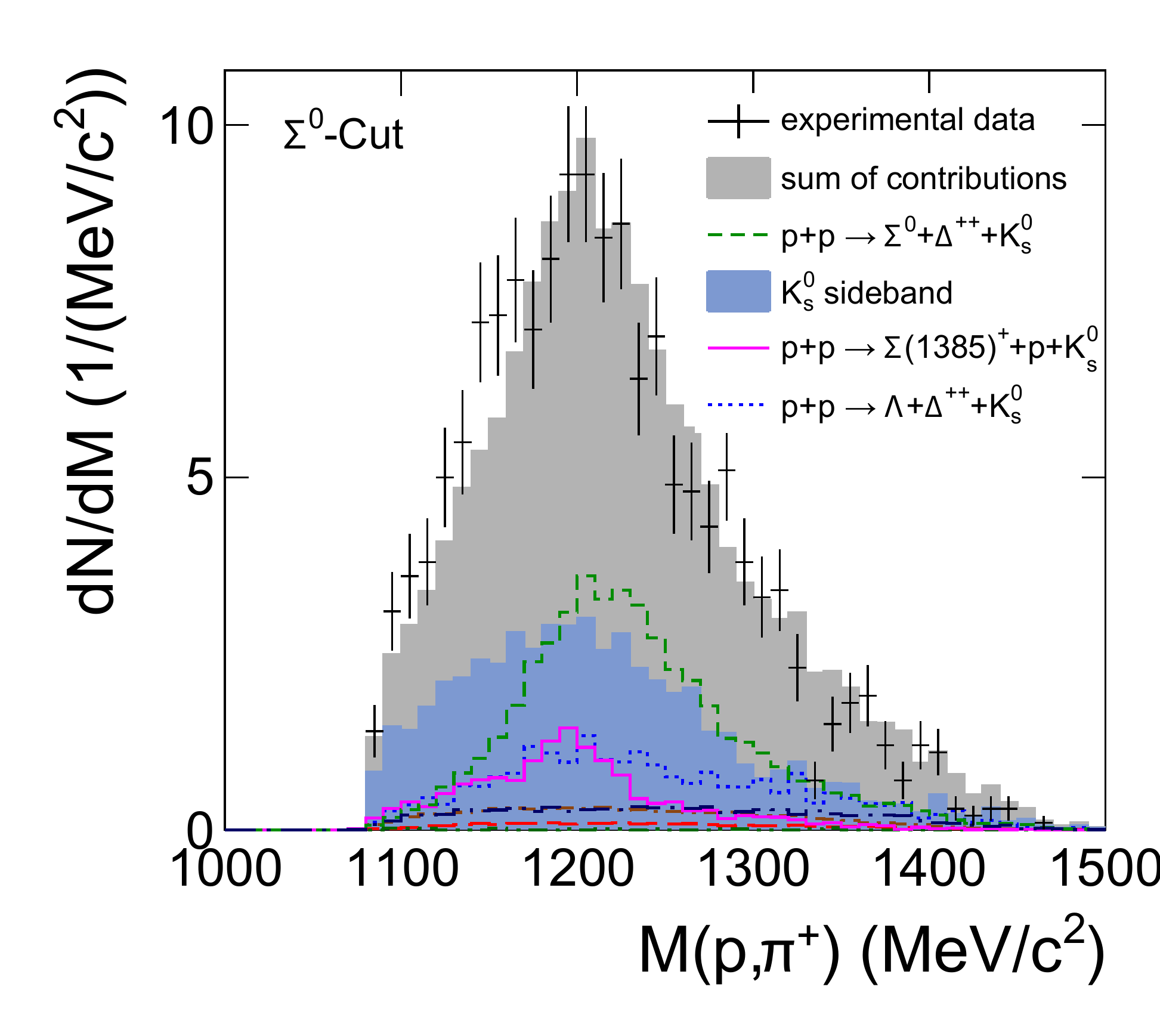}
\caption{\label{fig:pp_ExclIMS} Color online. $\mathrm{p\pi^+}$-invariant mass distribution after the $\Sigma^0$-cut on the MM(p,$\pi^+,\pi^+,\pi^-$) distribution (Fig.~\ref{fig:pp_ExclMMABCD}) and with a cut on the $\mathrm{K^0_S}$ mass in the $\pi^+\pi^-$-invariant mass spectrum (Fig.~\ref{fig:pp_ExclInvMassK0s}). The gray histogram corresponds to the sum of simulated contributions plus the background defined by the sideband sample. The same color code and line styles are used as in Figs.~\ref{fig:pp_ExclMMABCD},~\ref{fig:pp_ExclMMABD} and \ref{fig:pp_ExclMMABCD_hhcutMMABD}.}
\end{figure}

To determine the cross sections of the different $\mathrm{K^0}$ production channels under the assumption that no 
interferences occur between the different channels, a simultaneous fit over five observables (three missing mass and two invariant mass distributions) has been carried out. The experimental data have been fitted with the simulated cocktail described above together with the sideband sample. In addition to the two missing mass spectra MM(p,$\pi^+$,$\pi^+$,$\pi^-$) and MM(p,$\pi^+$,$\pi^-$) (Figs.~\ref{fig:pp_ExclMMABCD} and \ref{fig:pp_ExclMMABD}) three other distributions were included in the simultaneous fit. The first is the missing mass distribution MM(p,$\pi^+$,$\pi^+$,$\pi^-$)$_{CUT}$ obtained after the selection on MM(p,$\pi^+,\pi^-) >\, 1270\, \mathrm{MeV/c^2}$ (Fig. \ref{fig:pp_ExclMMABCD_hhcutMMABD}). As illustrated in Fig.~\ref{fig:pp_ExclMMABD} with this selection one gets rid of the contribution from the channel $\mathrm{p+p \rightarrow \Sigma^+ + p + K^0_S}$, so that the fit has a stronger constraint for the sideband contribution below $1000\,\mathrm{MeV/c^2}$ in the MM(p,$\pi^+,\pi^+,\pi^-$) 
spectrum. The last two spectra that have been included in the fit are the p$\pi^+$-invariant mass spectra (M$\mathrm{(p,\pi^+)}_{\Lambda}$ and M$\mathrm{(p,\pi^+)}_{\Sigma^0}$) obtained with a cut on the $\Lambda$ and $\Sigma^0$ peaks in the MM(p,$\pi^+,\pi^+,\pi^-$) distribution, respectively. The selected regions are marked with arrows in Fig. \ref{fig:pp_ExclMMABCD}. The obtained M$\mathrm{(p,\pi^+)}_{\Lambda}$ and M$\mathrm{(p,\pi^+)}_{\Sigma^0}$ distributions are shown in Figs.~\ref{fig:pp_ExclIML} and \ref{fig:pp_ExclIMS}. These distributions contain direct information about a possible contribution of a $\Delta^{++}$ decaying into proton and $\pi^+$ and allow to distinguish between the reactions $\mathrm{p+p \rightarrow \Lambda + \Delta^{++} + K^0_S}$ and $\mathrm{p+p \rightarrow \Sigma^0 + \Delta^{++} + K^0_S}$. 

The start parameters chosen for the minimization process are the cross sections $\mathrm{\sigma^{fit}_{ch}}$ quoted in Table~\ref{tab:pp_K0Cocktail_Table} normalized to the elastic p+p cross section as follows for each reaction: 
\begin{equation}
\mathrm{F^{start}_{ch} = \frac{\sigma^{fit}_{ch}}{N^{sim}_{ch}} \cdot \frac{N^{tot}_{el}}{\sigma_{el}}}.
\end{equation}
Here, $\mathrm{N^{sim}_{ch}}$ corresponds to the number of events simulated for each channel. $\mathrm{N^{tot}_{el}}$ is the total number of elastic events, whereas $\mathrm{\sigma_{el}}$ is the cross section for elastic events in the analyzed data set. Both numbers have been determined in a separate analysis \cite{Rustamov:2010zz}. The cross sections $\mathrm{\sigma^{fit}_{ch}}$ have been estimated from a fit to measured cross sections at various beam energies using the cross section parametrization from Eq. 34 in \cite{Sibirtsev:1998dh} and are listed in Table~\ref{tab:pp_K0Cocktail_Table}.

One has to mention here that the cross sections quoted for the channel $\mathrm{p+p \rightarrow \Lambda + p + \pi^{+} + K^{0}}$ have been determined without taking into account that the proton and the $\pi^{+}$ might stem from a $\Delta^{++}$ decay. To limit the free parameters in the fitting procedure, the cross sections of the reactions belonging to class C1 and C2 have been summed up within each class and treated each as one contribution. As explained before, this is valid because the shape in the fitted variables are similar. In addition the contribution of the sideband sample was allowed to vary within $\pm$30\% to compensate the fact that the yield of the sideband sample was determined on the basis of a small data sample - namely on the sample, which fulfills the condition MM(p,$\pi^+,\pi^-) <\, 1100\,\mathrm{MeV/c^2}$. Finally, the simultaneous fit delivers a scaling factor $\mathrm{f_{ch}}$ for each channel such that the following condition is fulfilled:
\begin{equation}
\mathrm{N_{exp} = \sum_{ch} f_{ch} \cdot F^{start}_{ch}}.
\end{equation}
It has not been mentioned yet that the reaction cross sections may depend on the angle of one of the produced particles. Such a dependence has little impact on the $\mathrm{\chi^2/NDF}$ and the p$\pi^+$-invariant mass spectra, but will affect the extracted reaction cross sections, as will be shown below. Hence, two measured angular distributions have been included and a method has been developed to determine the strength of the anisotropy for the two $\mathrm{\Delta^{++}}$ reactions. In the simulation the $\mathrm{\Sigma^++p+K^0}$ final states have been weighted such that the angular distribution of the $\mathrm{K^0}$'s reproduces the one reported in \cite{AbdelBary:2012vw} which has been extracted for $\mathrm{p_{beam} = 3059\,MeV/c}$. The angular distribution for the channel $\mathrm{p+p\rightarrow\Sigma(1385)^++n+K^+}$ has been extracted from \cite{Agakishiev:2011qw} and employed to weight the reaction $\mathrm{p+p\rightarrow\Sigma(1385)^++p+K^0}$.

\begin{figure}[h]
\begin{minipage}[b]{0.9\linewidth}
\includegraphics[width=\textwidth]{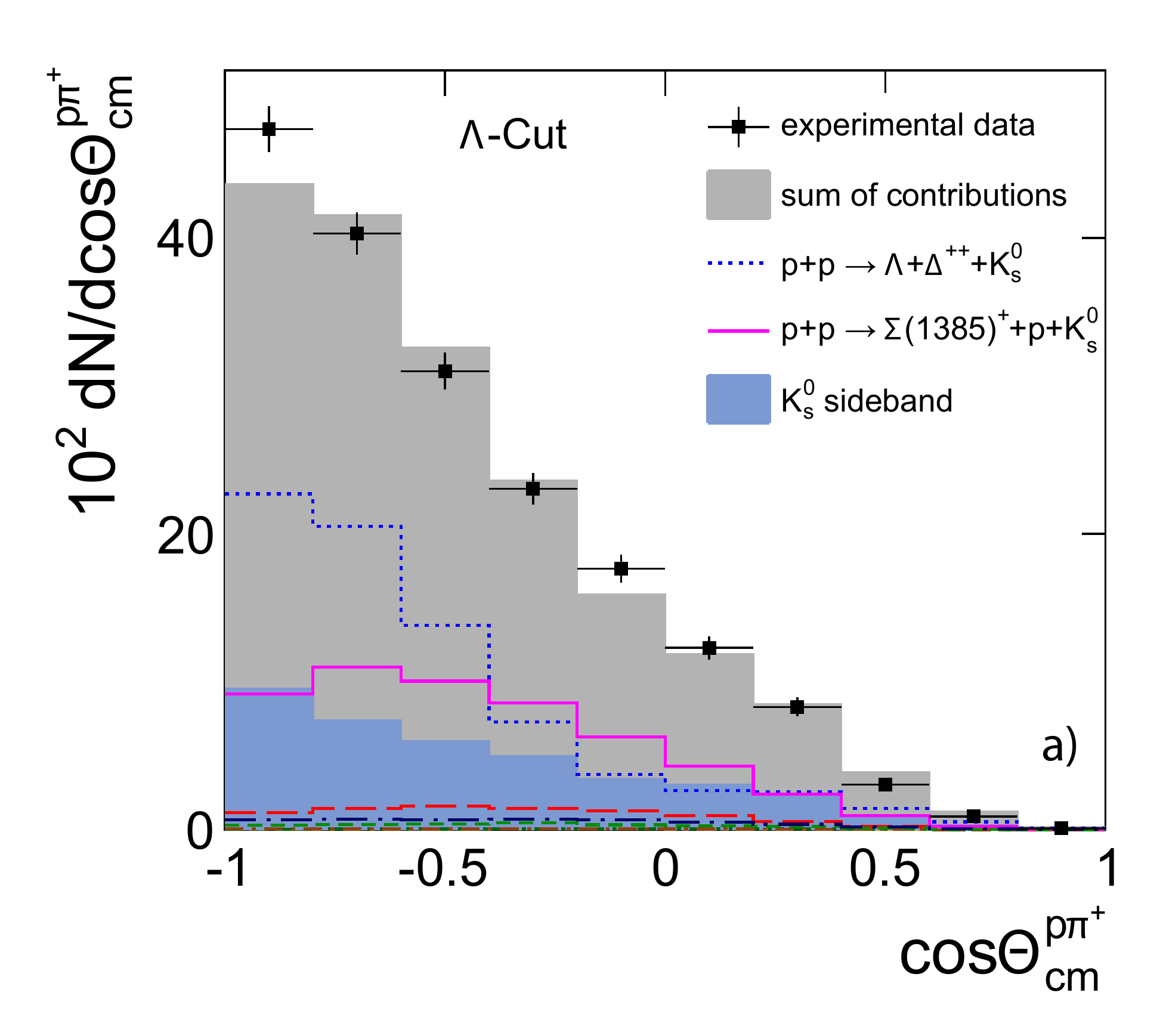}
\end{minipage}
\begin{minipage}[b]{0.9\linewidth}
\includegraphics[width=\textwidth]{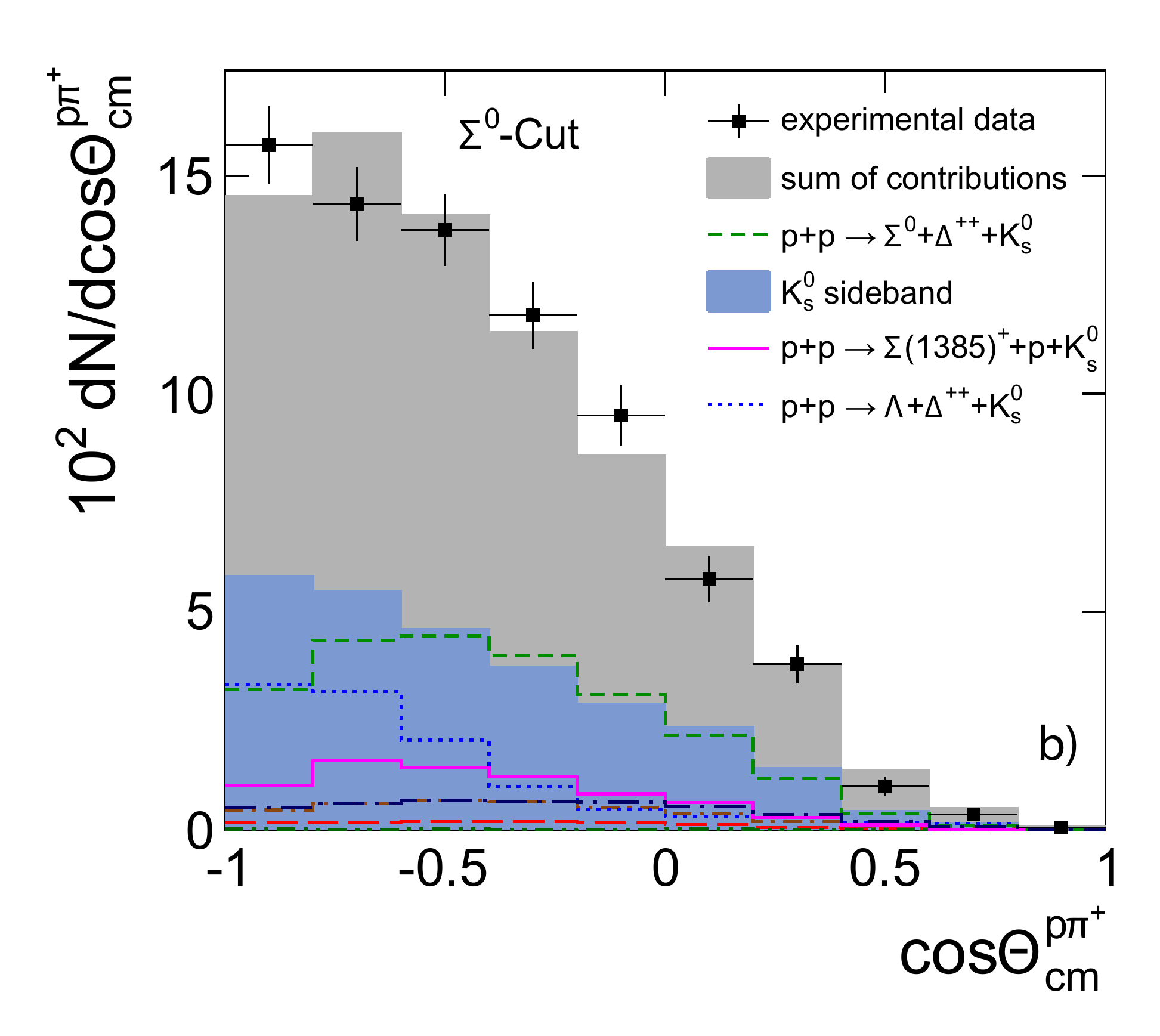}
\end{minipage}
\caption{\label{fig:pp_ExclCosThDPP} Color online. Angular distributions of p$\pi^+$-pairs in the center of mass reference system either in the $\Lambda$-cut (panel a)) or in the $\Sigma^0$-cut (panel b)) with a cut on the $\mathrm{K^0_S}$ mass in the $\pi^+\pi^-$-invariant mass spectrum (Fig.~\ref{fig:pp_ExclInvMassK0s}). The gray histogram corresponds to the sum of simulated contributions plus the background defined by the sideband sample. The same color code and line styles are used as in Figs.~\ref{fig:pp_ExclMMABCD},~\ref{fig:pp_ExclMMABD} and \ref{fig:pp_ExclMMABCD_hhcutMMABD}.}
\end{figure}

As far as the channels $\mathrm{p+p \rightarrow \Lambda + \Delta^{++} + K^{0}}$ and $\mathrm{p+p \rightarrow \Sigma^{0} + \Delta^{++} + K^{0}}$ are concerned, the angular anisotropy of the $\mathrm{\Delta^{++}}$ has been directly extracted from this data set. By looking at the $\mathrm{\cos{\Theta^{p\pi^+}_{cm}}}$ distributions in Fig.~\ref{fig:pp_ExclCosThDPP}, where panels a) and b) correspond to the $\mathrm{\Lambda}$ and the $\mathrm{\Sigma^0}$ selection, respectively, it becomes clear, that an anisotropic $\mathrm{\Delta^{++}}$ production is needed to reproduce the experimental data in both spectra. Without anisotropy in these two reactions the $\mathrm{\chi^{2}/NDF}$ for both angular distributions together would be only 25.64. By tuning the angular anisotropies of these two spectra, we have been able to extract the strength of the anisotropy. The angular distribution in the CMS has been parametrized using the Legendre polynomial function in Eq.~\ref{equ:LegendrePol}.
\begin{equation}\label{equ:LegendrePol}
\begin{split}
\mathrm{\frac{d\sigma}{dcos\Theta_{cm}}}\, =\, &\mathrm{A_0 + A_1cos\Theta_{cm} + A_2\frac{1}{2}(3cos^2\Theta_{cm} - 1)}\\
&\mathrm{+ A_4\frac{1}{8}(35cos^4\Theta_{cm} - 30cos^2\Theta_{cm} + 3)}
\end{split}
\end{equation}

\begin{figure}[h]
\includegraphics[width=0.45\textwidth]{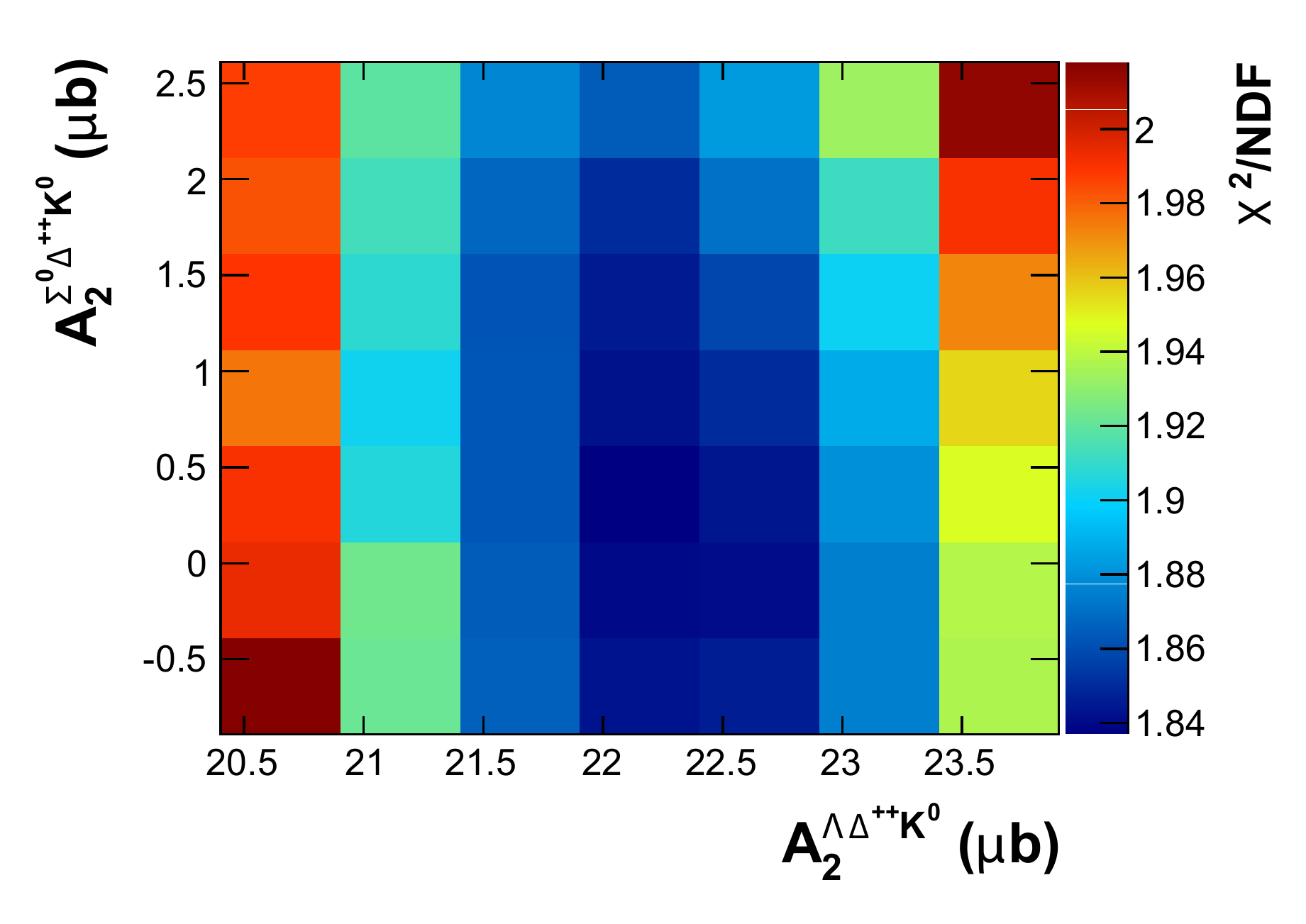}
\caption{\label{fig:pp_Chi2DPPAngle} Color online. $\mathrm{\chi^2/NDF}$ distribution in the $\mathrm{\cos{\Theta^{p\pi^{+}}_{cm}}}$ spectra in Fig.~\ref{fig:pp_ExclCosThDPP} for the variation of the second coefficients $\mathrm{A_2}$ of the Legendre polynomials of the channels $\mathrm{p+p \rightarrow \Lambda + \Delta^{++} + K^{0}}$ and $\mathrm{p+p \rightarrow \Sigma^{0} + \Delta^{++} + K^{0}}$.}
\end{figure}

The second coefficient of the polynomial representing the anisotropy has been varied in the range $\mathrm{A_2 = 20.65...23.65\,\mu b}$ for the channel $\mathrm{\Lambda + \Delta^{++} + K^{0}}$ and $\mathrm{A_2 = -0.64...2.36\,\mu b}$ for the reaction $\mathrm{\Sigma^{0} + \Delta^{++} + K^{0}}$. The coefficient $\mathrm{A_0}$ related to the yield of a reaction has been kept constant at 14.84$\mathrm{\,\mu b}$ and 4.63$\mathrm{\,\mu b}$ respectively. For each pair of $\mathrm{A_2}$ coefficients, the simultaneous fit over the five observables has been repeated. With the resulting scaling factors for each simulated contribution, the $\mathrm{\chi^{2}/NDF}$ in both $\mathrm{\cos{\Theta^{p\pi^+}_{cm}}}$ spectra could be determined. A plot of the $\mathrm{\chi^{2}/NDF}$ distributions for every pair of $\mathrm{A_2}$ coefficients can be found in Fig.~\ref{fig:pp_Chi2DPPAngle}. Finally, the combination which delivered the minimum $\mathrm{\chi^{2}/NDF}$ of 1.827 has been selected. This corresponds to $\mathrm{A_2^{\
Lambda\Delta^{++}K^{0}} = 22.15\,\mu b}$ and  $\mathrm{A_2^{\Sigma^0\Delta^{++}K^{0}} = 0.36\,\mu b}$. This means that the $\mathrm{\Delta^{++}}$ associated with a $\mathrm{\Lambda}$ production requires a rather strong anisotropy, whereas the production together with a $\mathrm{\Sigma^0}$ behaves almost like phase-space. A similar observation was made in the study of the 3-body reactions $\mathrm{p+p \rightarrow p + K^+ + \Lambda/\Sigma^0}$ at $\mathrm{p_{beam} = 3059\,MeV/c}$ by the COSY-TOF collaboration, where an opposite behavior of the angular distributions was found depending on the hyperon content \cite{AbdelBary:2012vw}. This points to different production mechanisms involving intermediate resonances to create either a $\mathrm{\Lambda}$K-pair or a $\mathrm{\Sigma^0}$K-pair. In the first case only $\mathrm{N^*}$ resonances are possible, while the production of a $\mathrm{\Sigma^0}$K-pair can also occur via $\mathrm{\Delta^*}$ resonances. 
Coming back to the simulation used in this analysis, the coefficients of all Legendre polynomials employed for the four anisotropic reactions are listed in Table~\ref{tab:pp_ExclLegPolCoef}. All other channels have been simulated isotropically in phase-space. 
\begin{figure}[h]
\begin{minipage}[b]{0.9\linewidth}
\includegraphics[width=\textwidth]{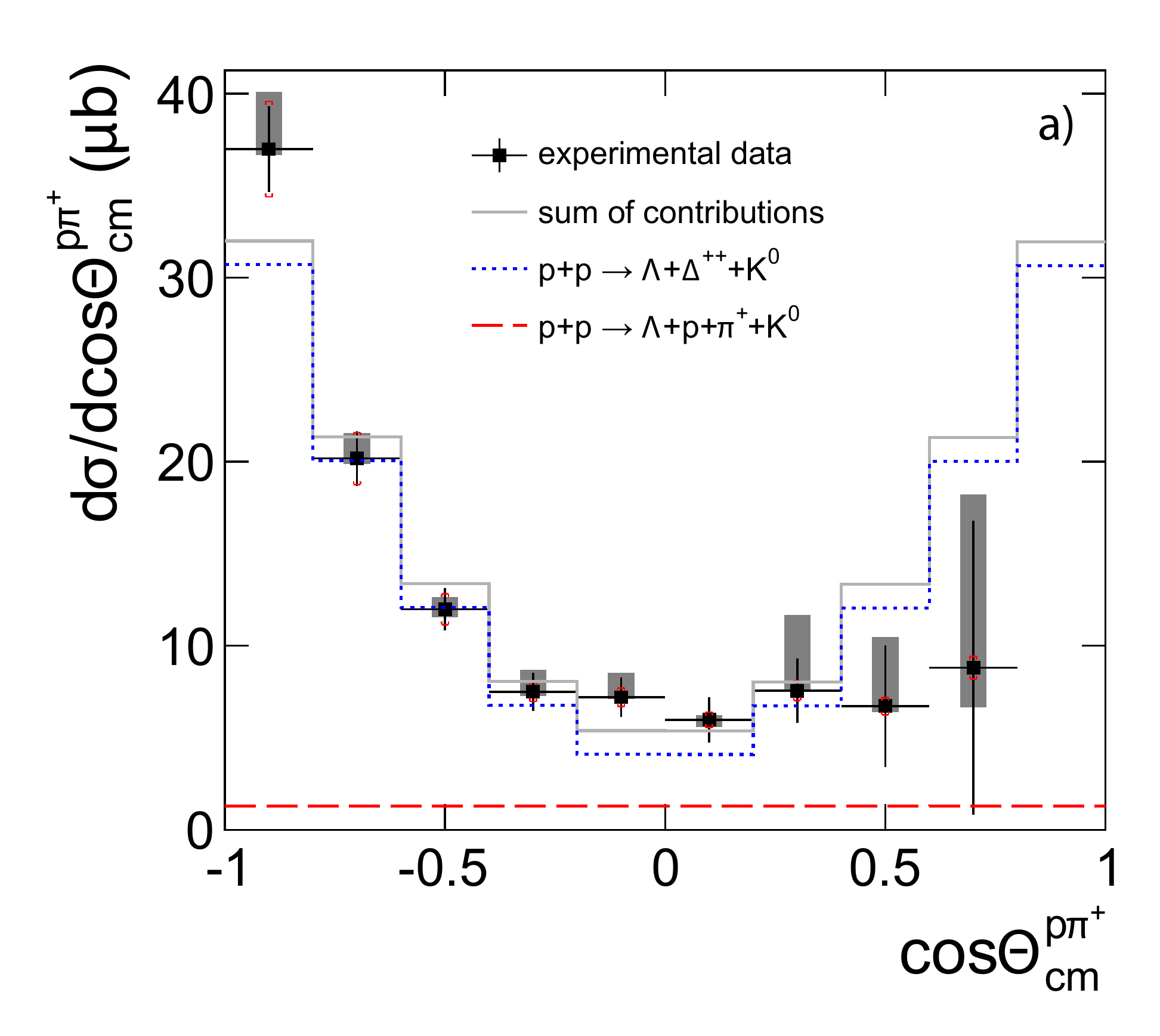}
\end{minipage}
\begin{minipage}[b]{0.9\linewidth}
\includegraphics[width=\textwidth]{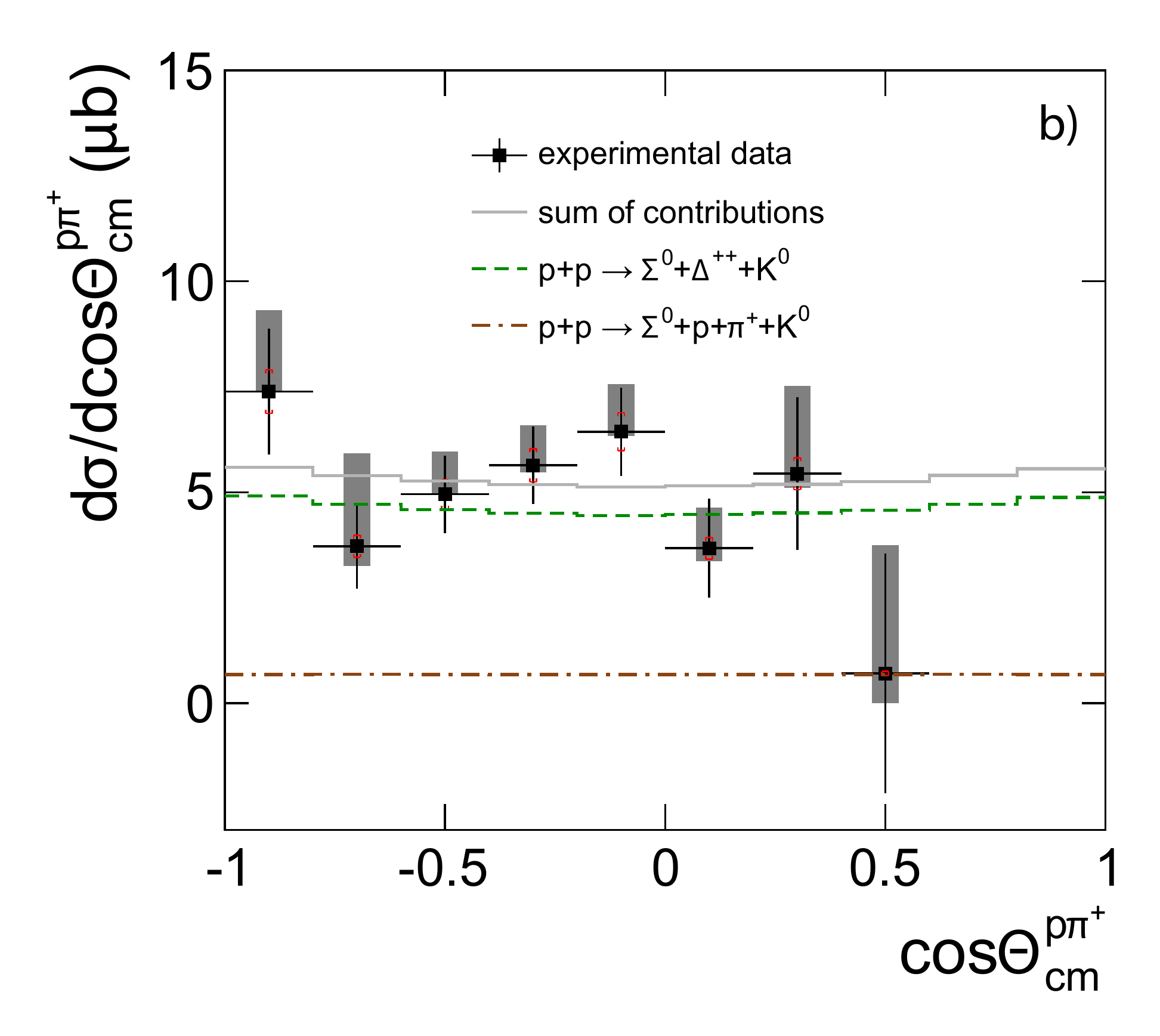}
\end{minipage}
\caption{\label{fig:pp_ExclCosThDPP_corr} Color online. Absolute normalized angular distributions of p$\pi^+$-pairs in the center of mass reference system corrected for acceptance and efficiency. Panel a) includes contributions from the reactions $\mathrm{p+p \to \Lambda + (p + \pi^+)/\Delta^{++} + K^0}$ and panel b) from the reactions $\mathrm{p+p \to \Sigma^0 + (p + \pi^+)/\Delta^{++} + K^0}$. The grey histogram corresponds to the sum of simulated contributions. The gray bands are the systematical uncertainties, whereas the red cups indicate the $7\,\%$ uncertainty from the absolute normalization to elastic scattering cross sections \cite{Rustamov:2010zz}.}
\end{figure}
\begin{table}[h]
\begin{center}
\caption{\label{tab:pp_ExclLegPolCoef} Coefficients of the Legendre polynomials included for the angular distributions of the listed particles in the corresponding reactions in [$\mu$b]. Coefficients of channel $\mathrm{p+p\rightarrow\Sigma^++p+K^0}$ are taken from \cite{AbdelBary:2012vw}. The coefficients for $\mathrm{p+p\rightarrow\Sigma(1385)^++p+K^0}$ was assumed to be the same as found in \cite{Agakishiev:2011qw} for the reaction $\mathrm{p+p\rightarrow\Sigma(1385)^++n+K^+}$.}
\begin{ruledtabular}
\begin{tabular}{llccc}
Reaction, $p+p\rightarrow$&cos&$A_0$&$A_1$&$A_2$\\
\colrule
$\Sigma^++p+K^0$&$\Theta^{K^0}_{cm}$&13.15&-0.40&4.37\\
$\Sigma(1385)^++p+K^0$&$\Theta^{\Sigma(1385)^+}_{cm}$&7.23&0.00&10.24\\
$\Lambda + \Delta^{++} + K^{0}$&$\Theta^{\Delta^{++}}_{cm}$&14.84&0.00&22.15\\
$\Sigma^{0} + \Delta^{++} + K^{0}$&$\Theta^{\Delta^{++}}_{cm}$&4.63&0.00&0.36\\
\end{tabular}
\end{ruledtabular}
\end{center}
\end{table}
\begin{table*}[htb]
\begin{center}
\caption{\label{tab:pp_ExclCrossSec} Cross sections of the exclusive $\mathrm{K^0}$ reactions. Here $\mathrm{\sigma_{anisotropic}}$ means, that the four channels listed in Table~\ref{tab:pp_ExclLegPolCoef} include an anisotropic angular distribution, while $\mathrm{\sigma_{isotropic}}$ means, that all channels were simulated isotropic. The first uncertainties correspond to statistical errors from the experimental data. The second uncertainties are the systematic errors from the variation of $\mathrm{K^0_S}$ secondary vertex cuts by $\pm$20\% and the variation of the integral in the $\pi^{+}\pi^{-}$-invariant mass distribution used for the sideband sample by again $\pm$20\%. The third uncertainties stem from the systematic uncertainties of the normalization to the elastic cross section \cite{Rustamov:2010zz}.} 
\begin{ruledtabular}
\begin{tabular}{lll}
$\mathrm{K^0}$ reactions&$\mathrm{\sigma_{isotropic}\,[\mu b]}$&$\mathrm{\sigma_{anisotropic}\,[\mu b]}$\\
\colrule
$p+p \rightarrow \Sigma^++p+K^0$&$24.25\pm0.63^{+2.42}_{-1.80}\pm1.70$&$26.27\pm0.64^{+2.57}_{-2.13}\pm1.84$\\
$p+p \rightarrow \Lambda+p+\pi^++K^0$&$2.37\pm0.02^{+0.18}_{-2.35}\pm0.17$&$2.57\pm0.02^{+0.21}_{-1.98}\pm0.18$\\
$p+p \rightarrow \Sigma^0+p+\pi^++K^0$&$1.40\pm0.02^{+0.41}_{-1.40}\pm0.10$&$1.35\pm0.02^{+0.10}_{-1.35}\pm0.09$\\
$p+p \rightarrow \Lambda+\Delta^{++}+K^0$&$25.56\pm0.08^{+1.85}_{-1.45}\pm1.79$&$29.45\pm0.08^{+1.67}_{-1.46}\pm2.06$\\
$p+p \rightarrow \Sigma^0+\Delta^{++}+K^0$&$9.17\pm0.05^{+1.45}_{-0.11}\pm0.64$&$9.26\pm0.05^{+1.41}_{-0.31}\pm0.65$\\
$p+p \rightarrow \Sigma(1385)^++p+K^0$&$13.15\pm0.05^{+1.91}_{-2.07}\pm0.92$&$14.35\pm0.05^{+1.79}_{-2.14}\pm1.00$\\
\end{tabular}
\end{ruledtabular}  
\end{center}
\end{table*}
\subsection{Results of the simultaneous fit}\label{sec:pp_Results}
Including the angular distributions and applying the minimization procedure as described above, we have achieved a global $\mathrm{\chi^2/NDF}$ of 2.57 in the five fitted variables from the simultaneous fit. The fit results are shown in the introduced missing mass, invariant mass and angular distributions, where the contributions of simulated reactions and the sideband sample are plotted (Figs.~\ref{fig:pp_ExclMMABCD},~\ref{fig:pp_ExclMMABD},~\ref{fig:pp_ExclMMABCD_hhcutMMABD}-\ref{fig:pp_ExclIMS},~\ref{fig:pp_ExclCosThDPP}). Only the spectra of multi-pion reactions (C2) are not shown, as the resulting contribution with 3.91$\mathrm{\cdot 10^{-7} \mu b}$ is very small. However, a good description of the experimental data by the fitted contributions can be observed. Especially from both p$\pi^+$-invariant mass  distributions it becomes clear, that a dominant contribution of the resonant channels ($\mathrm{\Lambda+\Delta^{++}+K^0}$ and $\mathrm{\Sigma^0+\Delta^{++}+K^0}$ to the $\mathrm{Y+p+\pi^++K^0}$ final 
state) is required to reproduce the shape of the data. In view of the single invariant mass spectra a $\mathrm{\chi^2/NDF}$ of 1.29 in the $\mathrm{\Lambda}$ and 2.44 in the $\mathrm{\Sigma^0}$ selection are achieved. If one would assume that the total yield of the final states $\mathrm{Y+p+\pi^++K^0}$ originates from non-resonant production, the $\mathrm{\chi^2/NDF}$ in these two spectra would degrade to 5.29 and 3.35 respectively.

Since the simulated $\mathrm{K^0}$ cocktail describes the discussed spectra fairly well together with the sideband sample, one can correct the angular distributions of the reactions $\mathrm{p+p \rightarrow Y+(p+\pi^+)/\Delta^{++}+K^0}$ for the HADES acceptance and efficiency. For this purpose, all modeled contributions have been subtracted from the experimental angular distributions (Figs.~\ref{fig:pp_ExclCosThDPP}) except the just mentioned reactions. Depending on the applied hyperon-selection in the angular distributions, the contribution of the other hyperon channels have been subtracted, too. Consequently, the angular distribution with the $\mathrm{\Lambda}$-cut should be left over with the reactions $\mathrm{\Lambda+(p+\pi^+)/\Delta^{++}+K^0}$ only and analogous in the case of the $\mathrm{\Sigma^0}$. The correction of these distributions has been performed with the simulations of the two resonant and the two non-resonant reactions by comparing the acceptance and efficiency filtered simulation to its 
initial distributions. As a result one can observe the clearly anisotropic production of the $\mathrm{p\pi^+}$-system if produced together with a $\mathrm{\Lambda}$ in Figure~\ref{fig:pp_ExclCosThDPP_corr} panel a). Panel b) displays an almost isotropic production of the $\mathrm{\Sigma^0+(p+\pi^+)/\Delta^{++}+K^0}$ final states. As already mentioned above, this can be seen as a consequence of the different productions mechanisms involving intermediate $\mathrm{N^*}$ and $\mathrm{\Delta^*}$ resonances to generate a $\mathrm{\Lambda}$K- or a $\mathrm{\Sigma}$K-pair, which at our energy is a common process.

To translate the known relative contributions of the simulated channels into cross sections, one has to scale the cross sections from the fitted cross section parameterizations to measured data ($\mathrm{\sigma^{fit}_{ch}}$) with the factor $\mathrm{f_{ch}}$, which was determined by the simultaneous fit of the reactions to the five experimental observables mentioned above.
\begin{equation}
\mathrm{\sigma^{tot}_{ch} = f_{ch} \cdot \sigma^{fit}_{ch}}.
\end{equation}
The total cross sections for all measured reactions are given in Table~\ref{tab:pp_ExclCrossSec} together with the cross sections assuming an isotropic simulation of all channels for comparison. In the Table $\mathrm{\sigma_{anisotropic}}$ means that four of the channels ($\mathrm{\Sigma^++p+K^0}$, $\mathrm{\Sigma(1385)^++p+K^0}$, $\mathrm{\Lambda+\Delta^{++}+K^0}$, $\mathrm{\Sigma^0+\Delta^{++}+K^0}$) include an anisotropic angular distribution, while $\mathrm{\sigma_{isotropic}}$ means that all channels have been simulated isotropically. The listed systematic uncertainties stem from a $\pm$20\% variation of each $\mathrm{K^0_S}$ secondary vertex cut and 
again a $\pm$20\% variation of the integral in the $\mathrm{\pi^+\pi^-}$-invariant mass distribution, that is used for the sideband sample to model the background.
From these measured cross sections one can conclude that the resonant ($\mathrm{\Delta^{++}}$) reaction is at least 10 times higher than the non-resonant channel in case of a production together with a $\mathrm{\Lambda}$. In case of a production together with a $\mathrm{\Sigma^0}$ it is at least a factor of 6 higher. For the reactions of class C2 one cannot make any conclusions about their finite cross sections, as their contribution to the selected events is too small. For the group of reactions classified as "other $\mathrm{K^0}$ reactions" (C3) a decrease of their summed cross sections by a factor of 0.82 has been extracted from this minimization. Comparing the extracted cross sections to the cross sections from a fitted parametrization \cite{Sibirtsev:1998dh} (Table~\ref{tab:pp_K0Cocktail_Table}) we see large differences. Especially the non-resonant reactions decrease significantly. That is because for the cross section parametrization also the resonant reactions leading to the same final states are 
included, as in most of the analysis it was not possible to distinguish them. Thus, only few measurements of the reaction $\mathrm{p+p \rightarrow \Lambda + \Delta^{++} + K^0}$ exist and even none of $\mathrm{p+p \rightarrow \Sigma^0 + \Delta^{++} + K^0}$, which implies a rather uncertain fit of the data, so that we observe a strong rise of its cross sections. The same situation is present for the reaction $\mathrm{p+p \rightarrow \Sigma(1385)^+ + p + K^0}$, where only one measurement with rather large uncertainties is taken into account.

\section{Summary and Conclusions}\label{sec:pp_SumCon}
Understanding the nucleon-nucleon reactions is of particular interest not only with respect to the plain NN interaction, but also with respect to particle (hadron) production. Still in the region of soft QCD the onset of multiple-hadron production implies a challenge for an appropriate phenomenological description. Due to the wide excitation spectrum of nucleons and Deltas, and also other baryons, the role of the individual states in particle production is an important issue for model building with the goal of first-principle formulations. Therefore, these strong-interaction aspects of hadron production and the mentioned properties are incorporated in transport models simulating for example proton-nucleon and nucleus-nucleus collisions, which need informations such as total reaction cross sections, angular distributions, scattering cross sections and others as an input. Especially at intermediate energies, where the models transit from resonance based to string fragmentation calculations, it is important to 
know the role of resonances.

We have reported on the exclusive analysis of proton-proton collisions at 3.5~GeV focusing on the 4-body reactions $p+p \rightarrow Y+p+\pi^{+}+K^{0}$ (Y = $\Lambda$, $\Sigma^{0}$). The proton and the $\pi^{+}$ in these reactions may be decay products of an intermediate $\Delta^{++}$ resonance. The data sample consisting of events with four charged particles (p, $\pi^+$, $\pi^+$, $\pi^-$) in the HADES acceptance includes additionally the reactions $\mathrm{p+p \rightarrow \Sigma^+ + p + K^0}$ and $\mathrm{p+p \rightarrow \Sigma(1385)^++ p + K^0}$. Besides these, a background contribution of non-strange channels remains after $\mathrm{K^0_S}$ preselection. This background has been modeled by a $\mathrm{K^0_S}$ sideband sample, which reproduces the distribution of the background events even in laboratory momenta and polar angles. Thus, it has allowed us to use this sample for the background description of the studied observables, which are first of all the missing mass distributions MM(p,$\pi^+,\pi^-$), MM(p,$\pi^+,\pi^+,\pi^-$) and MM(p,$\pi^+,\pi^+,\pi^-)_{CUT}$. As clear peaks at the $\Lambda$ and the $\Sigma^0$ masses show up in the missing mass distribution MM(p,$\pi^+,\pi^+,\pi^-$), we have used this information to classify the reaction $\mathrm{p+p \rightarrow Y+p+\pi^{+}+K^{0}}$ by their hyperon content. Indeed by studying the invariant mass distributions M(p,$\pi^+$) separately for the $\Lambda$ and the $\Sigma^0$ case, we have been also able to distinguish the production of the $\Delta^{++}$ resonance from the non-resonant production. An incoherent set of 14 $\mathrm{K^0}$ production channels has been simulated via the Pluto event generator. Together with the defined sideband sample this cocktail has been simultaneously fitted to the mentioned three missing mass and two invariant mass spectra, which has allowed to determine the relative contributions of the involved channels. The measured angular anisotropies for the reactions $\mathrm{p+p \rightarrow \Sigma^+ + p + K^0}$ and $\mathrm{p+p \rightarrow \Sigma(1385)^++ p + K^0}$ have been included for this process from \cite{AbdelBary:2012vw,Agakishiev:2011qw}. Furthermore, an iterative method has been developed using the two angular distributions $\mathrm{\cos{\Theta^{p\pi^+}_{cm}}}$ for the $\Lambda$ and $\Sigma^0$ region to extract the strengths of the anisotropy that occur for the reactions $p+p \rightarrow Y+\Delta^{++}+K^{0}$. It has been found that the $\Delta^{++}$ in the reaction associated with the $\Lambda$ hyperon is produced in a quite anisotropic way, whereas in the generation together with a $\Sigma^0$ it is emitted almost isotropically. The inclusion of these angular anisotropies has some effect on the obtained reaction cross sections, that can be found in Table~\ref{tab:pp_ExclCrossSec}. We can conclude that at intermediate energies resonances play a dominant role in the formation of hadrons. The results show a 6-10 times higher resonance contribution compared to non-resonant reactions. From theoretical point of view, the resonance model as 
described in \cite{Tsushima:1998jz} incorporates a somewhat consistent treatment for the formation of p$\pi^+$-pairs as found in this analysis, as there all these pairs are generated through the intermediate $\Delta^{++}$ resonance. Nevertheless, the model overestimates the cross sections of these reactions and angular distributions and channels including $\Sigma(1385)$ are missing in this approach. In the end, the reported results need to be consistently implemented and reproduced by the models in order to deliver more precise interpretations of nucleon-nucleus and heavy ion reactions, especially for the upcoming experiments of HADES and CBM at the FAIR facility that will perform measurements in an energy range from 2-50~AGeV \cite{Friman:2011zz,Fortov:2012tw}.

\begin{acknowledgments}
The authors are grateful to J. Aichelin and E. Bratkovskaya for the stimulating discussions.
The HADES collaboration gratefully acknowledges the support by the grants LIP Coimbra, Coimbra (Portugal) PTDC/FIS/113339/2009, SIP JUC Cracow, Cracow6 (Poland): N N202 286038 28-JAN-2010 NN202198639 01-OCT-2010, HZ Dresden-Rossendorf (HZDR), Dresden (Germany) BMBF 06DR9059D, TU M\"unchen, Garching (Germany) MLL M\"unchen: DFG EClust 153, VH-NG-330 BMBF 06MT9156 TP5 GSI TMKrue 1012 NPI AS CR, Rez, Rez (Czech Republic) MSMT LC07050 GAASCR IAA100480803, USC - S. de Compostela, Santiago de Compostela (Spain) CPAN:CSD2007-00042, Goethe University, Frankfurt (Germany): HA216/EMMI HIC for FAIR (LOEWE) BMBF:06FY9100I GSI F\&E.
\end{acknowledgments}

\bibliography{K0_pp_excl}

\end{document}